\documentclass[iop,revtex4]{emulateapj}
\usepackage{apjfonts}
\usepackage{lscape}

\usepackage[french, english]{babel} %support francais et anglais

\usepackage{longtable}
\usepackage{amsmath}
\usepackage{natbib}
\usepackage{url}

\newcommand{\Msun}{\mbox{$M_{\odot}$}}

\newcommand {\apgt} {\ {\raise-.5ex\hbox{$\buildrel>\over\sim$}}\ }
\newcommand {\aplt} {\ {\raise-.5ex\hbox{$\buildrel<\over\sim$}}\ } 
\newcommand{\kms}{\hbox{km~s$^{-1}$}}
\newcommand{\vsini}{\hbox{$v\sin i$}}
\newcommand{\vrad}{\hbox{RV}}

\slugcomment{Accepted to ApJ}

\begin{document}

\title{BANYAN. III. Radial velocity, Rotation and X-ray emission of low-mass star candidates in nearby 
young kinematic groups}

\author{Lison Malo\footnote{Based on observations obtained at the Canada-France-Hawaii Telescope (CFHT) 
which is operated by the National Research Council of Canada, the Institut National des Sciences de 
l''Univers of the Centre National de la Recherche Scientique of France, and the University of Hawaii.}, 
\'Etienne Artigau, Ren\'e Doyon, David Lafreni\`ere,
Lo\"{i}c Albert and Jonathan Gagn\'e}
\affil{D\'epartement de physique and Observatoire du Mont-M\'egantic, 
Universit\'e de Montr\'eal, Montr\'eal, QC H3C 3J7, Canada}

\email{malo@astro.umontreal.ca, doyon@astro.umontreal.ca} 

\begin{abstract}

Based on high-resolution spectra obtained with PHOENIX at Gemini-South, CRIRES at
VLT-UT1, and ESPaDOnS at CFHT, we present new measurements of the radial and
projected rotational velocities of 219 low-mass stars.
The target likely membership was initially established using the Bayesian 
analysis tool recently presented in \citet{2013malo}, taking into account only the position, 
proper motion and photometry of the stars to assess their membership probability. 
In the present study, we include radial velocity as an additional input to our analysis, 
and in doing so we confirm the high membership probability for 130 candidates: 
27 in $\beta$ Pictoris, 22 in Tucana-Horologium, 25 in Columba, 7 in Carina, 
18 in Argus and 18 in AB Doradus and 13 with an ambiguous membership. 
Our analysis also confirms the membership of 57 stars proposed in the literature.
A subsample of 16 candidates was observed at three or more epochs, allowing
us to discover 6 new spectroscopic binaries. 
The fraction of binaries in our sample is 25\%, consistent with values 
in the literature. 
Twenty percent of the stars in our sample show projected rotational velocities (\vsini) higher 
than 30 \kms\ and therefore are considered as fast rotators. 
A parallax and other youth indicators are still needed to fully confirm the 
130 highly probable candidates identified here as new {\it bona fide} members. 
Finally, based on the X-ray emission of {\it bona fide} and highly probable group members, we 
show that for low-mass stars in the 12-120 Myr age range, the X-ray luminosity is an excellent 
indicator of youth and better than the more traditionally used R$_{\rm X}$ parameter, the ratio
of X-ray to bolometric luminosity.

\end{abstract}

\keywords{Galaxy: solar neighborhood ---
Methods: statistical --- 
Stars: distances, kinematics, low-mass, moving groups, pre-main sequence ---
Techniques: radial velocities, spectroscopy }

%%%%%%%%%%%%%%%%%%%%%%%%%%%%%%%%%%%%%%%%%%%%%%%%%%%%%%%%%%%%%%%%%%%%%%%%%%%%%%%%%%%%%%%%%
\section{Introduction}
%%%%%%%%%%%%%%%%%%%%%%%%%%%%%%%%%%%%%%%%%%%%%%%%%%%%%%%%%%%%%%%%%%%%%%%%%%%%%%%%%%%%%%%%%

The search for young low-mass stars in the solar neighborhood has gained tremendous 
momentum recently as the youth and intrinsic faintness of these objects make them prime 
targets for planet searches through high-contrast imaging. 
This is both because young planetary-mass companions shine brighter and because 
low-mass stars are less luminous, which loosens the imaging contrast requirements. 
Finding more Young Moving Group (YMG) members also has intrinsic interest: to refine 
these YMG characteristics (initial mass function, age, velocities, spatial extent), 
to complete the solar neighborhood census, and to improve and develop new youth indicators.

As YMG members share common kinematics and span similar age ranges, one can search 
for overlooked members through a detailed kinematic and photometric analysis of nearby 
stars.
The Hipparcos mission, combined with multiple radial velocity studies 
(e.g., \citealt{2011anderson,2012debruijne}) led to the identification of the most massive
members of these groups, but the relatively shallow completeness magnitude ($V=7-9$)
prevented the detection of low-mass members (i.e., later than M5). 
Assuming that YMG stars follow a typical Initial Mass Function (IMF), the vast majority of 
low-mass members are expected to have gone unnoticed. 
This led to a number of efforts to
extend measurements of proper motion, radial velocity and parallax to the
nearby low-mass dwarfs (e.g., \citealt{2012shkolnik}).

Proper motion studies beyond the reach of $Hipparcos$ also benefitted greatly from the 
advent other all-sky surveys (e.g., \citealt{2009lepine,
2013zacharias}) reaching fainter magnitudes. 
In particular, infrared proper motions can now be obtained by correlating 
the WISE and 2MASS \citep{2012cutri} surveys, which is ideally suited to 
measure proper motions well into the substellar regime \citep[][]{2013rodriguez,2013agagne}.
These all-sky photometric surveys have now been supplemented by radial velocity (RV) measurements 
coming from various works. 
Surveys like the Radial Velocity Experiment \citep[RAVE; ][]{2006steinmetz, 2008zwitter} 
and SDSS \citep{2000york} performed RV measurements on a large sample of stars, while
others were focussed on old low-mass dwarfs \citep{1998delfosse,2009jenkins,2012chubak,2013soubiran}.
These RV studies are key for studying Galactic dynamics.
Finally, and not least, parallax measurements of solar neighborhood stars is spear-headed by the 
CTIOPI and RECONS group \citep[][]{2014riedel,2011riedel, 2010riedel} and other works 
\citep{2009lepine,2012shkolnik,2013liu} focusing on the lowest mass component of the solar neighborhood.

To complete the census of nearby young kinematic group members, kinematics alone is not 
enough to confirm membership, independent youth indicators are also mandatory to confirm the age 
of these stars. 
The high magnetic activity of young stars gives rise to various indicators such as 
$H\alpha$ emission, X-ray and UV \citep{2005preibisch, 2011shkolnik, 2011rodriguez, 2013rodriguez}.  
Other spectroscopic indicators include the Li abundance \citep{2008mentuch,2009dasilva} and others like 
sodium and calcium doublets (Na; 8183-8195\AA, Ca: 8498-8542\AA) that trace low-surface gravity 
\citep{2004lyo, 2006slesnick, 2008mentuch, 2009shkolnik}.
Stellar rotation and gyrochronology can also be used for constraining the age of an isolated 
star through a comparison with young stars in open clusters of known ages such as the Hyades, 
Pleiades and IC2391 \citep{2011irwin, 2009lopez, 2010messina,2010lopez}.  
Recently, \citet{2012areiners} presented the angular momentum evolution in low-mass stars, which 
can also be used as a youth indicator.
To assess the membership of a given star to the known YMGs, a complete description of these 
groups is essential: their space velocity, spatial extent, luminosity along with appropriate youth 
indicators.  
Based on such YMG descriptions, any given star candidate member of a YMG can have its observational
properties (proper motion, radial velocity, absolute magnitude)
compared against actual observations.

In the present paper, we present the follow-up of the 130 highly candidate members of young 
kinematic groups identified or re-identified in a previous paper 
\citep[][ hereafter Paper I]{2013malo}. 
In that paper, we presented an analysis tool (BANYAN\footnote{Bayesian Analysis for Nearby 
Young AssociatioNs: www.astro.umontreal.ca/$\sim$malo}) based on Bayesian inference to assess the membership 
probability of a given low-mass star to nearby YMGs.
This analysis has unveiled a large population of potentially new young low-mass stars but those 
stars cannot be confirmed as {\it bona fide} members of their respective association until 
knowledge of their radial velocity, parallax measurements and signs of youth are secured. 
The present work focuses on the radial and projected rotational velocity measurements 
which were measured through the cross-correlation technique using the infrared and optical domain. 
The paper is structured as follows.  
The sample and the observations are presented in $\S$~\ref{chap:deux} and data reduction is 
presented in $\S$~\ref{chap:trois}.
A presentation of the radial and projected rotational velocity measurements follows in 
$\S$~\ref{chap:quatre}. 
The identification and confirmation of highly candidate members is described in $\S$~\ref{chap:cinq} 
and $\S$~\ref{chap:six} is devoted to a discussion of rotation-age and X-ray luminosity relations.
Concluding remarks follow in $\S$~\ref{chap:sept}.

%%%%%%%%%%%%%%%%%%%%%%%%%%%%%%%%%%%%%%%%%%%%%%%%%%%%%%%%%%%%%%%%%%%%%%%%%%%%%%%%%%%%%%%%%
\section{Sample and Observations} \label{chap:deux}
%%%%%%%%%%%%%%%%%%%%%%%%%%%%%%%%%%%%%%%%%%%%%%%%%%%%%%%%%%%%%%%%%%%%%%%%%%%%%%%%%%%%%%%%%

A detailed account of our initial search sample has been presented in Paper I. 
In summary, our original sample of stars show chromospheric X-ray and H$\alpha$ emissions, and
have good I$_{c}$ photometry and proper motion measurements ($<$0.2mag and $>$4$\sigma$).
In Paper I, the $I$-band photometric data came from several studies. 
Recently, the UCAC4 \citep{2013zacharias} catalog published $I$-band photometric data from 
the APASS\footnote{www.aavso.org/apass} survey.
For this paper, we use an uniform source of $I$-band photometric data by using APASS-$i^{'}$ instead
of DENIS and SDSS-DR8 catalogs, when possible.
We transform the APASS-$i^{'}$ magnitude from UCAC4 
catalog to $I_{c}$ using conversion derived from the standards of Landolt \citep{2009landolt}.
The transformation is :
\begin{alignat}{2}
I_{c} &= i_{APASS} - 0.546
\end{alignat}
and accurate within 0.05 mag over 0.9$< I_{c}$ - $J <$2.0.

Using $I$-band photometry from the UCAC4 catalog has added 141 stars that were not included 
in Paper I because the $I$-band photometry was not available. 
Our initial sample was thus increased to 920 low-mass (K5V-M5V) stars, of which 
75 were previously identified as young stars in the literature.
By applying our Bayesian analysis to this extended sample, we found 247
candidate members with a membership probability ($P$) over 90\%, amongh which 50 were already 
proposed as candidate members in the literature. 
These candidates were restricted to the seven closest ($<$100\,pc) and youngest ($<$100\,Myr) 
co-moving groups considered in Paper I: TW Hydrae Association \citep[TWA;][]{1989delareza},  
$\beta$ Pictoris Moving Group \citep[$\beta$PMG;][]{2001bzuckerman}, 
Tucana-Horologium Association \citep[THA;][]{2000zuckermanwebb,2000torres},  
Columba Association \citep[COL;][]{2008torres}, 
Carina Association \citep[CAR;][]{2008torres},  Argus Association \citep[ARG;][]{2000makarov} 
and AB Doradus Moving Group \citep[ABDMG;][]{2004zuckerman}.

%%%%%%%%%%%%%%%%%%%%%%%%%%%%%%%%%%%%%%%%%%%%%%%%%%%%%%%%%%%%%%%%%%%%%%%%%%%%%%%%%%%%%%%%%
\subsection{Observations}
%%%%%%%%%%%%%%%%%%%%%%%%%%%%%%%%%%%%%%%%%%%%%%%%%%%%%%%%%%%%%%%%%%%%%%%%%%%%%%%%%%%%%%%%%

As described in Paper I, our statistical analysis yields a membership probability, a prediction
of the radial velocity and the most probable statistical distance for each star.
Membership confirmation requires both a parallax and RV measurements consistent with prediction
of the analysis.
Some candidate members listed in Paper I present ambiguous memberships i.e. they could belong to 
two or more YMGs; the radial velocities predicted for those groups differ by a few \kms\,.
Therefore, high resolution spectroscopy with accuracy of a few \kms\ was required to clarify the
membership of ambiguous candidates.
Observations were performed using two near-infrared spectrographs, 
PHOENIX \citep{2003hinkle} and CRIRES \citep{2004kaeufl}, and one optical spectrograph, ESPaDOnS \citep{2006donati}. 
In all cases, a set of slowly rotating radial-velocity standards were observed to calibrate our 
measurements. These template observations are summarized in Table~\ref{tab:template}. 

%%%%%%%%%%%%%%%%%%%%%%%%%%%%%%%%%%%%%%%%%%%%%%%%%%%%%%%%%%%%%%%%%%%%%%%%%%%%%%%%%%%%%%%%%
\subsubsection{PHOENIX}
%%%%%%%%%%%%%%%%%%%%%%%%%%%%%%%%%%%%%%%%%%%%%%%%%%%%%%%%%%%%%%%%%%%%%%%%%%%%%%%%%%%%%%%%%

In 2009 and 2010\footnote{Gemini ID program: GS-2009A-Q-89, GS-2009B-Q-45, GS-2010A-Q-32, 
GS-2010B-Q-18, QS-2010B-Q-89}, 
PHOENIX at Gemini-South was used with the 0.34$\arcsec$ wide slit in combination 
with the H6420 order-sorting filter (1.547 $\mu$m - 1.568 $\mu$m) to reach a resolving 
power of R $\sim$ 52,000 (4~pixels). 
This wavelength domain was selected as it is advantageously free of strong terrestrial 
absorption lines \citep{2002mazeh}. 

Observations were obtained with a typical ABBA dither pattern along the slit with individual 
exposures of 60 to 300 seconds depending on the target brightness.
This observing strategy facilitates the removal of sky emission lines through the subtraction 
of two consecutive images.
Along with each science target observation, a slowly radial velocity standard was observed 
immediately before or after.
We also observed 12 M0-M3V slow rotators, which were used to calibrate measurements of 
rotational broadening.
The typical signal-to-noise ratio (S$/$N) achieved was respectively 30-50 and 90-120 per 
spectral pixel for science targets and radial velocity standards.
Using that setup, 243 observations of 155 stars were obtained with a S$/$N sufficient to extract 
radial velocity measurements with an accuracy better than 1\,\kms. %Chiffe verifier 13/08/30

We also used two archival radial velocity standards (GJ~382 and GJ~447) from the Gemini archive, 
which were observed during the GS-2002B-Q-11 program with the same setup. 
The reduction was done using the same method as explained above.

%%%%%%%%%%%%%%%%%%%%%%%%%%%%%%%%%%%%%%%%%%%%%%%%%%%%%%%%%%%%%%%%%%%%%%%%%%%%%%%%%%%%%%%%%
\subsubsection{CRIRES}
%%%%%%%%%%%%%%%%%%%%%%%%%%%%%%%%%%%%%%%%%%%%%%%%%%%%%%%%%%%%%%%%%%%%%%%%%%%%%%%%%%%%%%%%%

CRIRES was used at VLT-UT1\footnote{VLT Period: 087.D-0510, 088.D-0553, 089.D-0592, 091.D-0641} 
with the 0.4$\arcsec$ and 0.2$\arcsec$ wide slit in the order 36 (1.555 $\mu$m), resulting in 
a resolving power of 50,000 and 80,000, respectively. %resolution 1 verifier 13/09/20
The dithering strategy and individual exposure times were similar to that used with PHOENIX.
Each science program also included observations of radial velocity standards and slow rotators. 
The achieved S$/$N was respectively 30-70 and 90-120 per spectral pixel
for science targets and radial velocity standards.
This observing strategy produced high S$/$N spectra allowing radial velocity measurements 
with accuracies better than 1\,\kms\ for 105 observations of 88 stars. %Chiffre verifier 13/08/30

%%%%%%%%%%%%%%%%%%%%%%%%%%%%%%%%%%%%%%%%%%%%%%%%%%%%%%%%%%%%%%%%%%%%%%%%%%%%%%%%%%%%%%%%%
\subsubsection{ESPaDOnS}
%%%%%%%%%%%%%%%%%%%%%%%%%%%%%%%%%%%%%%%%%%%%%%%%%%%%%%%%%%%%%%%%%%%%%%%%%%%%%%%%%%%%%%%%%

Starting in 2010, 53 stars were observed with ESPaDOnS, a visible light echelle spectrograph 
at CFHT\footnote{CFHT program: 11AC13, 11BC08, 11BC99, 12AC23, 12BC24, 13AC23, 13BC33}.
Observations were done using the "star+sky" mode combined with the "slow" and "normal" CCD 
readout mode, to get a resolving power of R$\sim$68,000 covering the 3700 to 10500 \AA 
spectral domain over 40 grating orders.
The total integration time per target was between 5 and 120 minutes. %Chiffre verifier 13/08/30

\begin{deluxetable}{lrrrrr}
\tabletypesize{\scriptsize}
\tablewidth{0pt}
\tablecolumns{6}
\tablecaption{Properties of radial velocity standards \label{tab:template}}
\tablehead{
\colhead{Name} & \colhead{Spectral} & \colhead{RV} & \colhead{$\vsini$} & \colhead{Ref.} & \colhead{Instrument}\\
\colhead{of template} & \colhead{Type} & \colhead{(km s$^{-1}$)} & \colhead{(km s$^{-1}$)} & \colhead{} & \colhead{}
}
\startdata
HIP 107345 & M1V & 2.3$\pm$0.5 & 8.2$\pm$0.1 & 1,1 & PH,CR \\
HIP 1993 & M0Ve & 6.4$\pm$0.1 & 7.3$\pm$2.3 & 1,1 & PH,CR \\
HIP 23309 & M0Ve & 19.4$\pm$0.3 & 5.8$\pm$0.3 & 1,1 & PH,CR \\
GJ 806 & M3V & -24.7$\pm$0.1 & 1.5  & 5,2 & ESP \\
GJ 908 & M2V & -71.3$\pm$0.1 & 3.0 & 5,2 & ESP \\
GJ 382 & M2V & 7.9$\pm$0.1 & $<$1.8 & 3,4 & CR,PH,ESP \\
GJ 447 & M4V & -31.1$\pm$0.1 & $<$2.5 & 3,4 & PH \\
GJ 729 & M3Ve & -10.5$\pm$0.2 & 2.5  & 5,1 & CR \\
HIP 67155 & M2V & 15.8$\pm$0.1 & $<$1.4 & 3,2 & CR,ESP \\
HIP 65859 & M0.5V & 14.56$\pm$0.1 & $<$2.0 & 3,2 & ESP \\
HIP 68469 & M1V & -25.8$\pm$0.1 & $<$2.0 & 3,2 & ESP
\enddata
\tablerefs{(1) \citealp{2006torres}; (2) \citealp{2009jenkins}; (3) \citealp{2002nidever}; (4) \citealp{2012breiners}; (5) \citealp{2012debruijne}}
\tablecomments{Instrument: PHOENIX (PH), CRIRES (CR) and ESPaDOnS (ESP)}
\end{deluxetable}

%%%%%%%%%%%%%%%%%%%%%%%%%%%%%%%%%%%%%%%%%%%%%%%%%%%%%%%%%%%%%%%%%%%%%%%%%%%%%%%%%%%%%%%%%
\section{Data reduction} \label{chap:trois}
%%%%%%%%%%%%%%%%%%%%%%%%%%%%%%%%%%%%%%%%%%%%%%%%%%%%%%%%%%%%%%%%%%%%%%%%%%%%%%%%%%%%%%%%%

%%%%%%%%%%%%%%%%%%%%%%%%%%%%%%%%%%%%%%%%%%%%%%%%%%%%%%%%%%%%%%%%%%%%%%%%%%%%%%%%%%%%%%%%%
\subsection{Near-infrared Data}
%%%%%%%%%%%%%%%%%%%%%%%%%%%%%%%%%%%%%%%%%%%%%%%%%%%%%%%%%%%%%%%%%%%%%%%%%%%%%%%%%%%%%%%%%

The data were reduced in a standard fashion for longslit spectroscopy, using custom $IDL$ routines. 
The individual nods frames were first dark subtracted and divided by a normalized flat field
(only for CRIRES data).
For PHOENIX data, flat image contains fringe thus we do not use the flat images.
The nods frames were then pair-subtracted, and we forced that dispersion to be perpendicular to 
the trace by correcting the slit inclinaison.
The 1-D spectrum extraction was obtained from the intensity difference image, where a median Gaussian fitting 
in the spatial direction was used (in both positive and negative traces).
Both the datasets were wavelength-calibrated using five strong OH emission lines \citep[1.5539711, 1.5540945, 
1.5545890, 1.5546393, 1.5570159 $\mu$m; ][]{2000rousselot} that encompass the wavelength domain used.

The noise in each spectrum was determined by measuring the pixel-to-pixel scatter along 
the extracted 1-D spectrum. 
Each spectrum was corrected for the barycentric velocity component of the Earth using the $IDL$
function $baryvel$ which is available as part of the astrolib package \footnote{idlastro.gsfc.nasa.gov}.
For visual binary stars, we reduced each component by fitting two Gaussian profiles to the trace.

%%%%%%%%%%%%%%%%%%%%%%%%%%%%%%%%%%%%%%%%%%%%%%%%%%%%%%%%%%%%%%%%%%%%%%%%%%%%%%%%%%%%%%%%%
\subsection{ESPaDOnS Data}
%%%%%%%%%%%%%%%%%%%%%%%%%%%%%%%%%%%%%%%%%%%%%%%%%%%%%%%%%%%%%%%%%%%%%%%%%%%%%%%%%%%%%%%%%

The ESPaDOnS observations were reduced by CFHT using UPENA 1.0, an in-house
software that calls the Libre-ESpRIT pipeline \citep{1997donati}. 
In the present analysis, we used the processed spectra with the continuum normalized to 
1 and automatic wavelength correction from telluric lines.

%%%%%%%%%%%%%%%%%%%%%%%%%%%%%%%%%%%%%%%%%%%%%%%%%%%%%%%%%%%%%%%%%%%%%%%%%%%%%%%%%%%%%%%%%
\section{Results} \label{chap:quatre}
%%%%%%%%%%%%%%%%%%%%%%%%%%%%%%%%%%%%%%%%%%%%%%%%%%%%%%%%%%%%%%%%%%%%%%%%%%%%%%%%%%%%%%%%%

The extracted high-resolution spectra enabled measurement of the science target radial 
velocity from the observed Doppler shift, and the projected rotational velocity (\vsini) 
from the width of the absorption lines. 
All 403 observations were analysed using the procedure described in Section~\ref{chap:quatreun}.
Table~\ref{tab:indivi} summarizes all measurements.

In addition, it was possible to investigate the multiplicity fraction of these stars by 
classifying them into three categories: 1) visual binaries (using the acquisition images); 
2) double-line spectroscopic binaries (SB2); and 3) single-line spectroscopic binaries (SB1).

%%%%%%%%%%%%%%%%%%%%%%%%%%%%%%%%%%%%%%%%%%%%%%%%%%%%%%%%%%%%%%%%%%%%%%%%%%%%%%%%%%%%%%%%%
\subsection{Radial and projected rotational velocity measurements} \label{chap:quatreun}
%%%%%%%%%%%%%%%%%%%%%%%%%%%%%%%%%%%%%%%%%%%%%%%%%%%%%%%%%%%%%%%%%%%%%%%%%%%%%%%%%%%%%%%%%

The radial velocity (\vrad) and projected rotational velocity (\vsini) were measured through 
a cross-correlation between each individual spectrum and a template.
The template consisted in radial velocity standard (see Table~\ref{tab:template}) convolved with 
rotational profiles and artificially Doppler shifted.
All spectra were effectively re-sampled at equal steps of $\log \lambda$ to properly sample data
in equal velocity bins.
In artificially broadening standard star spectra, the rotational profiles defined in \citet{1992Gray} 
were adopted, or more specifically we used the $IDL$ $astrolib$ routine $lsf\_rotate$. 
The limb darkening was set to a typical value of 0.6. 
The \vrad\ and \vsini\, parameters were scanned between -150 and 150 \kms\ and between 1 and 120 \kms,
respectively. 
The maximum of the Cross-Correlation Function (CCF) was found by fitting a Gaussian function to 
the central peak. 
That yielded the best value of \vrad\ and \vsini\,. 
For each template, a $\chi^2$ minimization map in 2-D (\vrad\ and \vsini) was generated and the 
template producing the lowest $\chi^2$ was adopted which yielded the final \vrad\ and \vsini\, measurements.

The wavelengths across which the analysis was performed were the same for all near-infrared datasets 
(between 1.553 and 1.5575 $\mu$m). 
For the optical data, cross-correlation was performed on an order by order basis between 531 to
793 nm. 
Only parts of orders devoid of strong telluric absorption were used, these are orders: 26, 27, 28, 29, 
30, 31, 33, 35, 36, 37 and 42.

For double-line spectroscopic binary cases, the two peaks seen in the CCF map were both fitted by
a Gaussian function. 

\begin{figure}[!hbt]
\epsscale{1.0}
\plotone{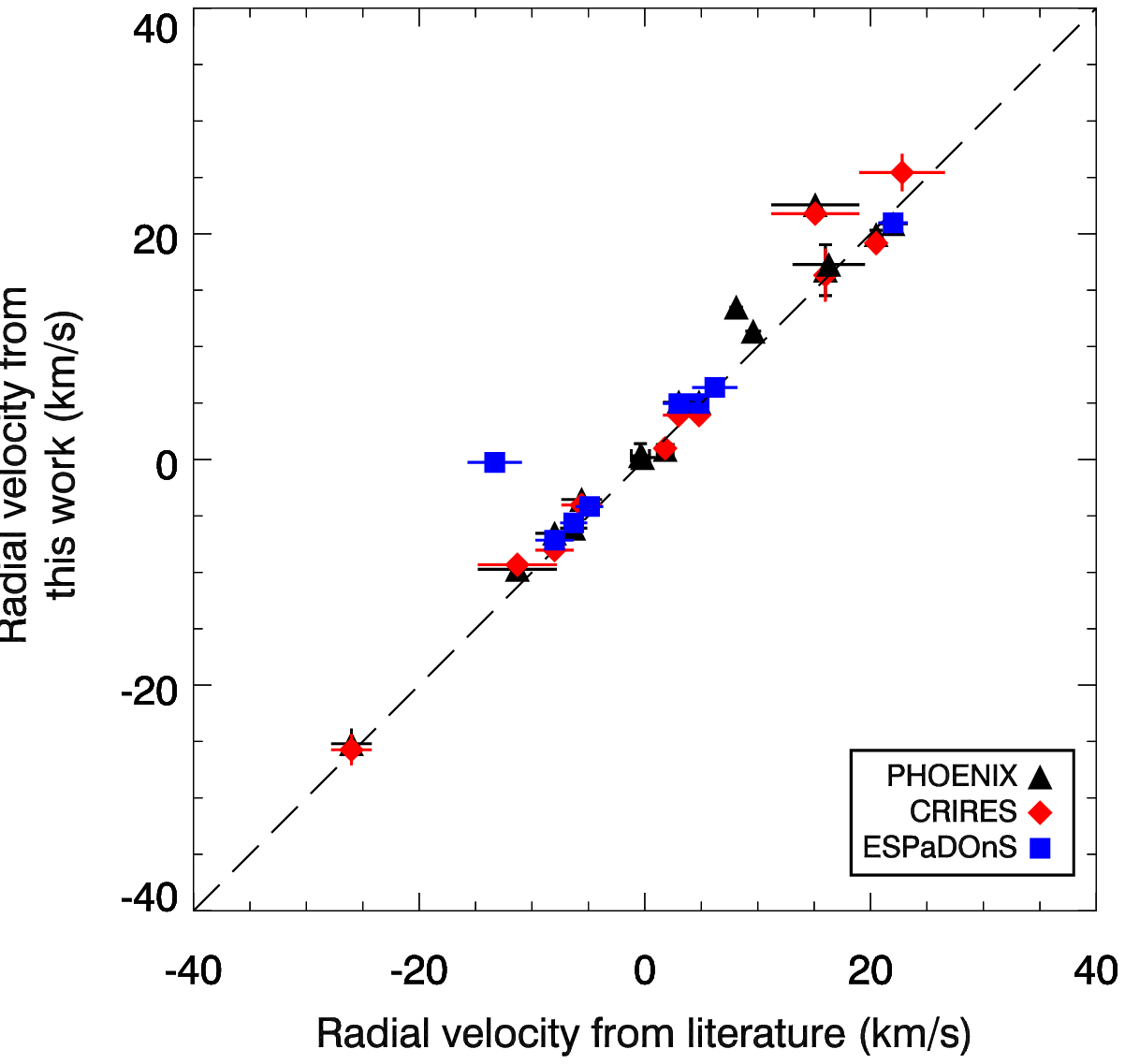}
\plotone{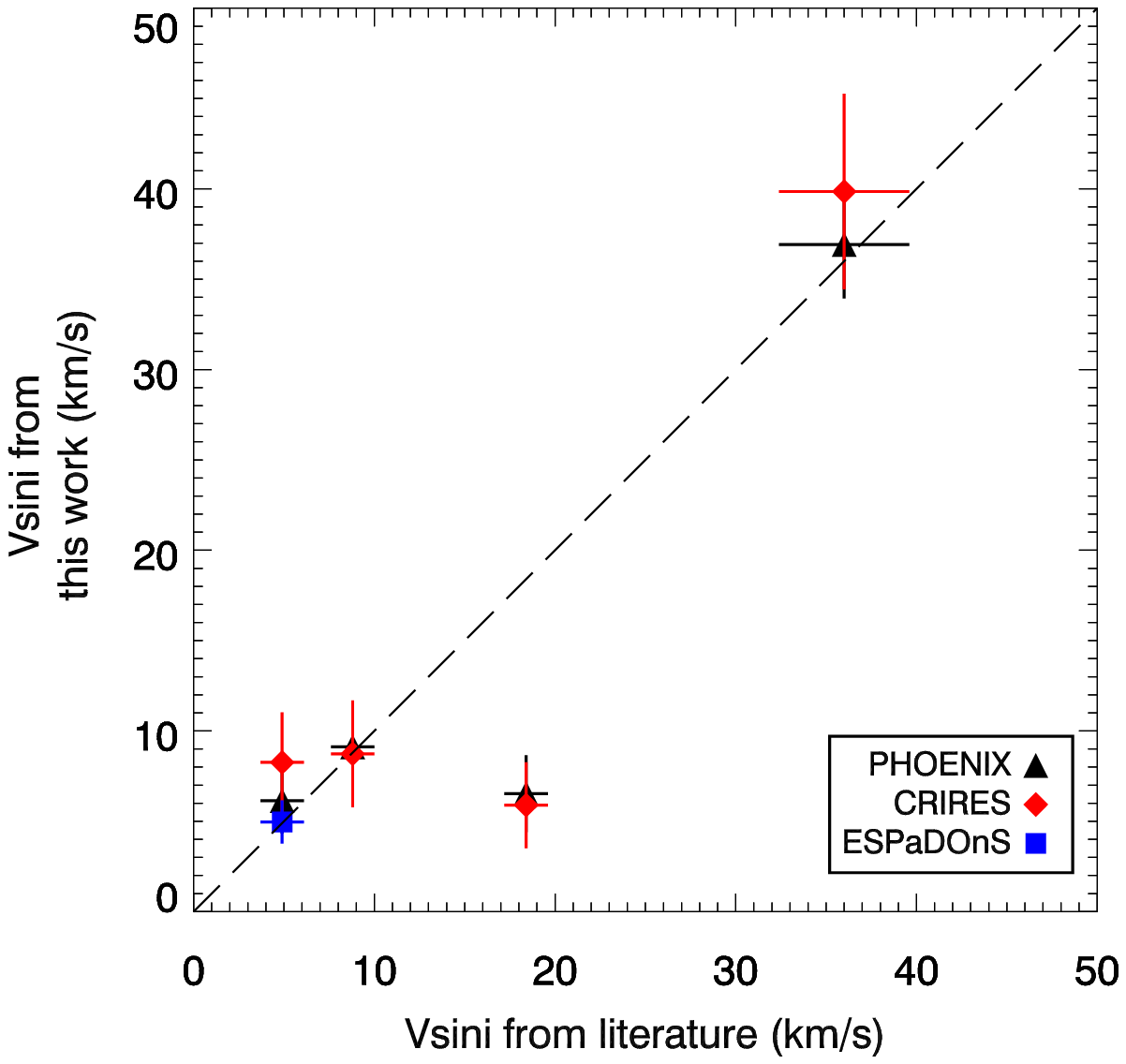}
\caption{\footnotesize{Radial and projected rotational velocity measurements for PHOENIX (black triangles), 
CRIRES (red diamonds) and ESPaDOnS (blue squares) compared to 
compiled measurements from the literature. The dotted line illustrates the one-to-one
relation between literature and measured velocities.} 
\label{fig:compare}}
\end{figure}

Figure~\ref{fig:compare} presents a comparison between radial and projected rotational velocity 
measurements from this analysis and those compiled from the literature. 
There is an excellent correlation between the literature values and those observed from this work, 
with a standard deviation of 1.3, 1.9 and 0.8 \kms\ for PHOENIX, CRIRES and ESPaDOnS, respectively. 
Except for one star (J05332558-5117131), we have similar results for projected rotational velocity 
measurements.
We cannot exclude that this exception is a spectroscopic binary, which was unresolved during our 
observations but mostly resolved during \citet{2006torres} observation.

%%%%%%%%%%%%%%%%%%%%%%%%%%%%%%%%%%%%%%%%%%%%%%%%%%%%%%%%%%%%%%%%%%%%%%%%%%%%%%%%%%%%%%%%%
\subsubsection{Uncertainty on radial and projected rotational velocity measurements} \label{chap:quatreunun}
%%%%%%%%%%%%%%%%%%%%%%%%%%%%%%%%%%%%%%%%%%%%%%%%%%%%%%%%%%%%%%%%%%%%%%%%%%%%%%%%%%%%%%%%%

\begin{figure}[!hbt]
\epsscale{1.2}
\plotone{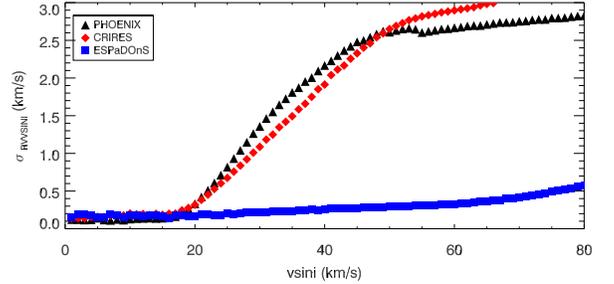}
\caption{\footnotesize{Systematic radial velocity offset induced by the convolution of a slowly 
rotating RV templates (GJ~382) as a function of \vsini. The CRIRES and PHOENIX datasets (red diamonds, black triangles) 
show a $>$ 1\,\kms\ systematic errors for \vsini\, values above $\sim$ 30 \kms, while ESPaDOnS data set
(order42: blue squares), which covers a much broader wavelength domain, is almost immune to this 
effect.} 
\label{fig:gj382}}
\end{figure}

Our sources of uncertainty on radial velocity measurement (similar for \vsini\, measurements) 
fall into three categories, which are expressed in the Equation~\ref{ensh}: \\

\begin{equation} \label{ensh}
\sigma_{RV}^{2} = \sigma_{noise}^{2} + \sigma _{rv std}^{2} + \sigma _{rv vsini}^{2}
\end{equation}

where $\sigma_{noise}$ is the statistical spectrum noise, $\sigma _{rv std}$ is the uncertainty of 
the RV standard and $\sigma _{rv vsini}$ is the systematic error as a function of \vsini.
The  $\sigma_{noise}$ was determined by repeating, through a 
Monte Carlo analysis, the procedure explained in Section~\ref{chap:quatre}, this time adding
Gaussian noise to each spectral pixel and measuring the standard deviation of the resulting distribution.
The second source of uncertainty taken into account is the error on RV and \vsini\, of the slowly-rotating 
radial velocity templates, whose values are tabulated in Table~\ref{tab:template}.
The third source is the systematic uncertainty from the \vrad\ offset which is effectively measured 
when comparing \vrad\ of an artificially broadened slowly-rotating template to its original self. 
That \vrad\ offset as a function of \vsini\, was determined for each template.
Figure~\ref{fig:gj382} presents the systematic \vrad\ offset as a function 
of \vsini\, when GJ~382 is used as a radial velocity template for the three instruments.
The correction is significant for \vsini\, $>$ 30 \kms\,. For example, a target spectrum correlated 
with that of GJ~382 with a \vsini\, of 30 \kms\ introduces a systematic \vrad\ offset of 
1.6 \kms, 1.1 \kms\ and 0.2 \kms\ with PHOENIX, CRIRES and ESPaDOnS, respectively. 

%%%%%%%%%%%%%%%%%%%%%%%%%%%%%%%%%%%%%%%%%%%%%%%%%%%%%%%%%%%%%%%%%%%%%%%%%%%%%%%%%%%%%%%%%
\subsection{Multiplicity fraction} \label{chap:quatredeux}
%%%%%%%%%%%%%%%%%%%%%%%%%%%%%%%%%%%%%%%%%%%%%%%%%%%%%%%%%%%%%%%%%%%%%%%%%%%%%%%%%%%%%%%%%

Properties of multiple systems provide strong empirical constraints on star formation
theories and evolution.
Moreover, double-lined spectroscopic binaries allow precise determination of dynamical 
properties of the components, including the mass ratio \citep{2010galvez-ortiz}.
One clear observational trend, which has gradually emerged is that the multiplicity 
fraction gradually decreases with decreasing stellar mass \citep{2012janson}.

During our follow-up program, a sample of hiterto unidentified visual or 
spectroscopic binaries was uncovered.
From the 219 observed stars, we found 37 visual binary (VB), 12 double-lined spectroscopic 
binaries (SB2) and 6 single-lined spectroscopic binaries (SB1) showing radial velocity 
variations of more than 5 \kms\,.  
Table~\ref{tab:binary} lists the properties of the binary systems found within our program.

It is worthwhile to note that for 78\% of the $\beta$PMG and ABDMG candidate binary systems,
our statistical analysis predicted an overluminosity compared to {\it bona fide} members. 
This overluminosity prediction decreases to 33\% for THA, COl, CAR and ARG candidates.
As stated in Paper I, this overluminosity could be the result of unresolved binarity.
For $\beta$PMG and ABDMG, our follow-up observations give credence to our analysis for
identifying binary systems.

We find a binary fraction for the low-mass components of 25\%, which is consistent with
the value published in the literature \citep[25\%,42\%][]{1997leinert,2004delfosse}.
\citet{2010shkolnik} find from a sample of 185 X-ray M dwarfs a spectroscopic binary 
fraction of 16\%.
\citet{2012janson} find a multiplicity fraction for M0V-M5V dwarfs and separations 
between 0.08$\arcsec$ and 6$\arcsec$ of 27$\pm$ 3\%.
\citet{2010bergfors} estimated the multiplicity fraction for young M0V-M6V dwarfs 
($<$600\,Myr) with an angular separation 0.1$\arcsec$ to 6$\arcsec$ to be 32$\pm$6\%.

We note that ultrafast rotators (\vsini\, $>$ 50\,\kms\,) could be unresolved spectroscopic binaries 
with two rapidly rotating components. 
Therefore, a second epoch of measurements is needed to rule out the binary status.
An example of such behaviour is the TWA candidate member SCR 1425-4113 \citep[see ][]{2014riedel}.
While the CRIRES observation did not enable us to detect the two binary components, we measured 
a \vsini\, of 95 \kms\,.
The second observation with ESPaDOnS enabled us to resolve the two components of the system.

\begin{deluxetable}{lccccc}
\tabletypesize{\scriptsize}
\tablewidth{0pt}
\tablecolumns{6}
\tablecaption{Properties of binary systems \label{tab:binary}}
\tablehead{
\colhead{Name} & \colhead{Sep.} & \colhead{q2/q1} & \colhead{Type of} & \colhead{Instr.} & \colhead{Refs.\tablenotemark{b}} \\
\colhead{(2MASS)} & \colhead{(arcsec)} & \colhead{(flux ratio)} & \colhead{binary\tablenotemark{a}} & \colhead{} & \colhead{}
}
\startdata
J00233468+2014282AB & \ldots & \ldots & VB & ESP & 1 \\
J00281434-3227556   & 0.6  & 0.71 & VB & PH & 1 \\
J00340843+2523498W  & \ldots & \ldots & VB & ESP & 5\\
J01132817-3821024   & 1.3   & 0.72 & VB & PH & 4 \\
J01535076-1459503   & 2.7   & 0.80 & VB & PH &  4\\
J04480066-5041255A  & \ldots & \ldots & VB & PH & 1 \\
J05045462-1415337   & 0.5  & 0.76 & VB & PH & 1 \\
J05100488-2340148   & 1.7  & 0.63 & VB & PH & 4 \\
J05100427-2340407A  & \ldots & \ldots & VB & PH & 4\\
J05130132-7027418   & 1.6   & 0.2  & VB & CR & 4 \\
J05142736-1514514   & 2.5   & 0.3  & VB & PH & 1 \\
J05241914-1601153AB & \ldots & \ldots & VB & PH & 6\\
J05254166-0909123AB & \ldots & \ldots & VB & PH & 7\\
J05301858-5358483   & 4.3  & 0.04 & VB & CR & 4 \\
J06002304-4401217   & 1.9  & 0.97 & VB & CR & 4 \\
J06153953-8433115   & 1.1  & 1.0  & VB & CR & 1 \\
J06434532-6424396AB & \ldots & \ldots & VB & PH,CR & 4\\
J07105990-5632596   & 0.9  & 0.35 & VB & PH & 1 \\
J07523324-6436308A  & \ldots & \ldots & VB & PH & 1 \\
J08224744-5726530   & 8.1   & 0.2  & VB & CR & 4 \\
J08412528-5736021   & 1.4  & 0.97 & VB & PH & 4 \\
J08472263-4959574   & 1.7  & 0.84 & VB & PH & 1 \\
J09032434-6348330   & 7.9   & 0.11 & VB & CR & 1 \\
J09423823-6229028   & 1.3  & 0.90 & VB & CR & 4 \\
J10120908-3124451   & 1.1 & 0.96 & VB & CR &  1,3\\
J11102788-3731520   & 1.4   & 0.70 & VB+SB2 & PH & 4\\
J14142141-1521215   & 1.3  & 0.34 & VB & PH & 1 \\
J16074132-1103073AB & \ldots & \ldots & VB & PH & 1 \\
J17104431-5300250A  & \ldots & \ldots & VB & PH & 1 \\
J17165072-3007104   & 1.6 & 0.92 & VB & CR & 1\\
J17243644-3152484   & 0.9 & 0.74 & VB & CR & 1\\
J17494867-4005431A  & \ldots & \ldots & VB & PH & 1\\
J18450097-1409053A  & \ldots & \ldots & VB & PH & 1 \\
J20434114-2433534   & 1.3  & 0.98 & VB & PH & 8 \\
J21103096-2710513   & \ldots & \ldots & VB & CR  & 1\\
J23002791-2618431   & 2.1  & 0.54 & VB & PH & 1 \\
J23204705-6723209   & 2.6  & 0.55 & VB & PH  & 1 \\
\cutinhead{Double-line binary }
J01351393-0712517 & \ldots & \ldots & SB2 & ESP &  1\\
J06131330-2742054 & \ldots & \ldots & SB2 & PH  & 3 \\ 
J07282116+3345127 & \ldots & \ldots & SB2 & ESP  & 9\\
J08185942-7239561 & \ldots & \ldots & SB2 & CR & 1\\
J08422284-8345248 & \ldots & \ldots & SB2 & PH,CR & 1\\
J08465879-7246588 & \ldots & \ldots & SB2 & PH  & 1 \\
J09361593+3731456 & \ldots & \ldots & SB2 & ESP & 5 \\
J14252913-4113323 & \ldots & \ldots & SB2 & ESP & 1,3 \\
J17462934-0842362 & \ldots & \ldots & SB2 & PH  & 1 \\
J18141047-3247344 & \ldots & \ldots & VB+SB2 & CR  & 10\\
J19420065-2104051 & \ldots & \ldots & SB2 & CR & 1 \\
J20100002-2801410 & \ldots & \ldots & SB2 & ESP  & 4\\
\cutinhead{Single-line binary }
J01132958-0738088 & \ldots & \ldots & SB1 & \ldots & 4\\
J02070176-4406380 & \ldots & \ldots & SB1 & \ldots & 1\\
J02485260-3404246 & \ldots & \ldots & SB1 & \ldots & 1\\
J15244849-4929473 & \ldots & \ldots & SB1 & \ldots & 1\\
J18495543-0134087 & \ldots & \ldots & SB1 & \ldots & 1\\
J22470872-6920447 & \ldots & \ldots & SB1 & \ldots & 1
\enddata
\tablenotetext{a}{Visual binary (VB), Single-line binary (SB1), Double-line binary (SB2). }
\tablenotetext{b}{(1) this work; (2) \citet{2006torres}; (3) \citet{2014riedel}; (4) \citet{2012janson}; (5) \citet{2010skiff}; (6) \citet{2010bergfors}; (7) \citet{2007daemgen}; (8) \citet{2009shkolnik}; (9) \citet{2010shkolnik}; (10) \citet{2010messina}.}
\end{deluxetable}

%%%%%%%%%%%%%%%%%%%%%%%%%%%%%%%%%%%%%%%%%%%%%%%%%%%%%%%%%%%%%%%%%%%%%%%%%%%%%%%%%%%%%%%%%
\section{Candidates Membership} \label{chap:cinq}
%%%%%%%%%%%%%%%%%%%%%%%%%%%%%%%%%%%%%%%%%%%%%%%%%%%%%%%%%%%%%%%%%%%%%%%%%%%%%%%%%%%%%%%%%

\begin{deluxetable*}{lrrrrrrr}
\tabletypesize{\scriptsize}
\tablewidth{0pt}
\tablecolumns{8}
\tablecaption{Main properties of young kinematic groups \label{tab:prop2}}
\tablehead{
\colhead{Name} & \colhead{$UVW$} & \colhead{$\sigma_{\sc UVW}$} & \colhead{$XYZ$} & \colhead{$\sigma_{\sc XYZ}$} & \colhead{Number} & \colhead{Age\tablenotemark{a}} & \colhead{Distance\tablenotemark{b}} \\
\colhead{of group} & \colhead{(km s$^{-1}$)} & \colhead{(km s$^{-1}$)} & \colhead{(pc)} & \colhead{(pc)} & \colhead{of objects} & \colhead{range (Myr)} & \colhead{range (pc)}
}
\startdata
TW Hydrae (TWA) & $-10.53,-18.27,-5.00$ & $3.50,1.17,2.15$ & $12.17,-43.23,21.90$ & $6.14,7.30,3.06$ & 12 & 8-12\tablenotemark{c} & 42-92 \\
$\beta$ Pictoris ($\beta$PMG) & $-11.16, -16.19, -9.27$ & $2.06, 1.32, 1.35$ & $4.35, -5.82, -13.29$ & $31.43, 15.04, 7.56$ & 44 & 12-22\tablenotemark{d} & 9-73 \\
Tucana-Horologium (THA) & $-9.93, -20.72, -0.89$ & $1.55,1.79,1.41$ & $11.80, -20.79,-35.68$ & $18.57,9.14,5.29$ & 42 & 20-40\tablenotemark{e} & 36-71\\
Columba (COL) & $-12.24,-21.27,-5.56$ & $1.08,1.22,0.94$ & $-28.22,-29.74,-28.07$ & $13.68,23.70,16.09$ & 20 & 20-40\tablenotemark{e} & 35-81\\
Carina (CAR) & $-10.50,	-22.36,-5.84$ & $0.99,0.55,0.14$ & $15.55,-58.53,-22.95$ & $5.66,16.69,2.74$ & 5 & 20-40\tablenotemark{e} & 46-88\\
Argus (ARG) & $-21.78,-12.08,-4.52$ & $1.32,1.97,0.50$ & $14.60,-24.67,-6.72$ & $18.60,19.06,11.43$ & 11 & 30-50\tablenotemark{f} & 8-68\\
AB Doradus (ABDMG) & $-7.11,-27.21,-13.82$ & $1.39,1.31,2.26$ & $-2.25,2.93,-15.42$ & $20.10,18.97,15.37$ & 48 & 70-120\tablenotemark{g} & 11-64\\
Field stars & $-10.92,-13.35,-6.79$ & $23.22,13.44,8.97$ & $-0.18,2.10,3.27$ & $53.29,51.29,50.70$ & 10094 &  & 3-150
\enddata
\tablenotetext{a}{Age range from the literature. Note that relative ages between moving groups are more accurate than absolute ones.}
\tablenotetext{b}{Members with published trigonometric distance only.}
\tablenotetext{c}{Age reference from \citet{1999webb, 2006barrado}}
\tablenotetext{d}{Age reference from $\beta$PMG \citet{2003song, 2007makarov}}
\tablenotetext{e}{Age reference from \citet{2001torres, 2000zuckermanwebb}}
\tablenotetext{f}{Age reference from \citet{2008torres, 2004barrado} }
\tablenotetext{g}{Age reference from \citet{2008mentuch, 2006lopez, 2005luhman} }
\end{deluxetable*}

The statistical analysis presented in Paper I assumed no prior knowledge of the radial velocity 
information for most systems\footnote{BANYAN web tool: www.astro.umontreal.ca/$\sim$malo}. 
The radial velocity obtained in this work can be used to further constraint their membership.  
Our analysis is based on a kinematical model that takes into account the mean Galactic space velocities ($UVW$) 
and Galactic positions ($XYZ$) along with their dispersion. 
Those parameters, given in Paper I, have been slightly modified to take into account several new proposed 
members from recent studies \citep[][]{2014riedel,2012shkolnik,2013weiberger,2011wahhaj}. 
The following stars are considered new {\it bona fide} members; in $\beta$PMG: 2MASSJ01112542+1526214
(LP~467-16AB), 2MASSJ05064946-2135038 (GJ~3332BC), 2MASSJ05064991-2135091 (GJ~3331A), HIP~23418ABCD 
and 2MASSJ03350208+2342356;
%ok2013-09-13 
in TWA: TWA2 and TWA12; and in ABDMG: 2MASSJ06091922-3549311 (CD-35 2722 AB), J07234358+2024588 (BD+20 1790) and HIP~107948.  
For the sole purpose of establishing the core membership of the association, two outlier stars were removed from the 
{\it bona fide} member list.
We removed HIP~51317 (M2V) from the ABDMG {\it bona fide} member list because its X-ray luminosity appears to be much 
more compatible with a field star (see Section~\ref{chap:sixdeux}).  
HIP~24947 in THA was also excluded because it has ambiguous radial velocity measurements \citep[15.2, 23.9 \kms;][]{2006bobylev,2007bobylev} that place it either in THA or COL.
Table~\ref{tab:prop2} summarizes the main properties of YMGs.

This study presents \vrad\ measurements for 368 low-mass stars, of which 202 are new measurements 
and 166 are compiled from the literature. 
We also presents \vsini\ measurements for 270 stars, of which 202 are new measurements 
and 68 are compiled from the literature
A weighted average of all \vrad\ and \vsini\, measurements was adopted for stars with multiple measurements (see Tables~\ref{tab:compilrv} and ~\ref{tab:compilvsini}).
Adding the RV information in our Bayesian tool, we found 130 highly-probable YMG
candidates with membership probabilities (P$_{v}$) in excess of 90\%. 
Of these, a subgroup of 117 can be uniquely associated with a single existing YMG, and the remaining 13 have 
high membership probabilities to two or more YMGs. 
A trigonometric distance is required to assign membership without ambiguity.
The 117 candidates are divided the following way between parent YMGs: 27 in $\beta$PMG,
22 in THA, 25 in COL, 7 in CAR, 18 in ARG and 18 in ABDMG. 
In addition, we confirmed the high membership probability of 57 candidates previously proposed in the literature. 
Table~\ref{tab:candprop} presents the properties of the strong YMG candidates which have a P$_{v}$ above 90\%. 

For fast rotators with \vsini\, $>$ 30 \kms\, the \vrad\ becomes more uncertain which prevents significant 
constraints on the membership to be set. 
However, the rapid rotation deduced from the large \vsini\, when combined with other youth indicators 
(Malo et al. in prep) help constrain the age within certain broad limits (see Section~\ref{chap:cinqun}). 
Candidates whose membership probability decreases when \vrad\ is included in the analysis but have a high
projected rotational velocity ($>$ 30 \kms\,) are tabulated in Table~\ref{tab:candprop}.

%%%%%%%%%%%%%%%%%%%%%%%%%%%%%%%%%%%%%%%%%%%%%%%%%%%%%%%%%%%%%%%%%%%%%%%%%%%%%%%%%%%%%%%%%
\subsection{Kinematic model prediction} \label{chap:cinqun}
%%%%%%%%%%%%%%%%%%%%%%%%%%%%%%%%%%%%%%%%%%%%%%%%%%%%%%%%%%%%%%%%%%%%%%%%%%%%%%%%%%%%%%%%%

\begin{figure}[!hbt]
\epsscale{1.0}
\plotone{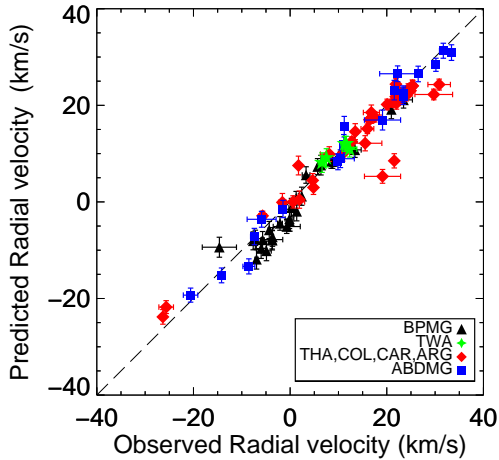}
\caption{\footnotesize{Comparison between predicted and observed radial velocities 
for the 111 candidate members (P$_{v}$ $>$ 90\% and excluding known binary).} \label{fig:fig11}}
\end{figure}

As explained in Section\,4 of Paper I, predicted radial velocity for candidate members are 
derived from a kinematic model, using Galactic space velocities of 
{\it bona fide} members as input parameters.
One way to assess the accuracy of this model is to look for systematic differences 
between predicted and observed radial velocity measurements.
Figure~\ref{fig:fig11} presents a comparison between predicted and observed radial velocities
for 111 candidate members with P$_{v}$ $>$ 90\%; as expected, the agreement between model
predictions and observations is excellent.

\begin{figure}[!hbt]
\epsscale{1.2}
\plotone{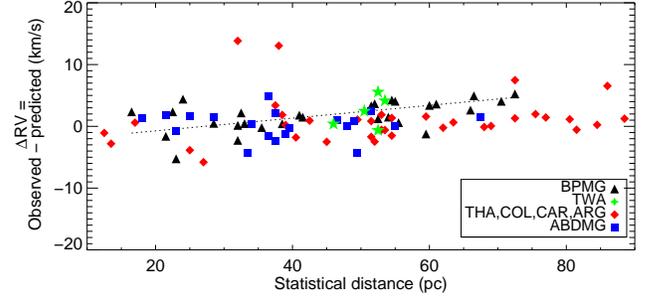}
\caption{\footnotesize{Comparison between predicted and observed radial velocities 
for YMGs as a function of the statistical distance.} \label{fig:fig12}}
\end{figure}

For $\beta$PMG candidate members, we observe a systematic difference of 3-5 \kms\ for a
radial velocity range between -10 and 0 \kms\,.
This systematic difference is also shown in Figure~\ref{fig:fig12} and seem to 
increase as a function of distance.
This may be a hint that the kinematic properties of low-mass members of $\beta$PMG differ
slightly from that of the bulk population.

%%%%%%%%%%%%%%%%%%%%%%%%%%%%%%%%%%%%%%%%%%%%%%%%%%%%%%%%%%%%%%%%%%%%%%%%%%%%%%%%%%%%%%%%%
\subsection{New bona fide members} \label{chap:cinqdeux}
%%%%%%%%%%%%%%%%%%%%%%%%%%%%%%%%%%%%%%%%%%%%%%%%%%%%%%%%%%%%%%%%%%%%%%%%%%%%%%%%%%%%%%%%%

Although our observations and analysis unveiled 117 new highly probable candidate members to YMGs,
a word of caution is necessary before assigning a firm membership to these objects.
These candidates will be attributed the status of {\it bona fide} members only after
their parallax is measured and shown to be consistent with the statistical distance predicted 
by our analysis. 
These candidates should also show evidence of youth through other indicators, e.g. lithium EW
strength, low-surface gravity, and - as argued in the next section - relatively strong X-ray 
luminosity.

Recent parallax and radial velocity measurements along with spectroscopic observations provide 
evidence for adding three new low-mass stars to the list of {\it bona fide} members. 
Those systems are briefly discussed below. 

2MASSJ00275035-3233238 (GJ~2006A) and 2MASSJ00275035-3233238 (GJ~2006B).
This binary system ($\rho$=18$\arcsec$, M3.5Ve+M3.5Ve) was first proposed to be a member
of $\beta$PMG by \citet{2014riedel} who measured a trigonometric distance of 32.3$\pm$1.8\,pc,
consistent with our predicted statistical distance of 32$\pm$2\,pc.
Its spectrum shows sign of low gravity (hence youth) as evidenced by the strength of the 
NaI8200 index.  
Our radial velocity measurements (8.8$\pm$0.2 \kms and 8.5$\pm$0.2 \kms) are consistent with 
the prediction (8.4$\pm$1.5 \kms) of our kinematic model and yield a very high
membership probability in $\beta$PMG : $P_{v}$ = 99.9\% and $P_{v+\pi}$ = 99.9\% . \\

2MASSJ06131330-2742054 (SCR 0613-2742AB).
This is another binary system ($\rho$=0.093$\arcsec$, M4V unresolved) proposed to be a member of 
$\beta$PMG by \citet{2014riedel} at a trigonometric distance of 29.4$\pm$0.9\,pc, consistent with 
our statistical distance of 25$\pm$6\,pc. 
Our RV (22.5$\pm$0.2~\kms) is also consistent with that measured by Riedel et al. (22.54$\pm$1.16 \kms), all of 
which yield a very high membership probability ($P_{v+\pi}$ = 99.9\%) in $\beta$PMG. 
The system also shows low-surface gravity based on the NaI8200 index.
There is no lithium absorption which is expected for $\beta$PMG {\it bona fide} members near 
the lithium depletion boundary \citep[][]{2014riedel,2010Yee} for $\beta$PMG.
It is interesting to note that our analysis correctly predicts the binary nature of this star 
without knowledge of the parallax and the RV.\\

2MASSJ14252913-4113323 (SCR 1425-4113AB).
This spectroscopic binary system (M2.5(sb)) was first proposed to be a potential member of 
TWA (with a marginal kinematic fit) by \citet{2014riedel}.
Its NaI8200 index is consistent with a low-gravity object. 
It shows EW Li absorption (595 m\AA) consistent with a membership in TWA. 
However, our radial velocity (-2.0$\pm$1.6 \kms) is more consistent with a membership in $\beta$PMG with a predicted
RV of -1.5$\pm$1.8 \kms compared to 1.5$\pm$2.8 \kms for TWA.
When constrained with the trigonometric distance (66.9$\pm$4.3~pc) measured by Riedel et al., our 
analysis yields a membership probability of 89.2\% in $\beta$PMG assuming a binary system.
This star is yet another system for which our analysis correctly predicts its binary 
nature without RV and parallax measurements.
If in $\beta$PMG, this star would be the most distant M dwarf of the $\beta$PMG. 
However, given its relatively large distance combined with a large Li EW much more consistent with an association 
younger than $\beta$PMG (e.g TWA),  one should be cautious before assigning a {\it bona fide} membership to $\beta$PMG 
for SCR 1425-4113AB. 
This is beyond the scope of this paper but one should consider other distant associations
\citep[e.g the Scorpius-Centaurus complex;][]{2012song} in our Bayesian  analysis to firmly 
establish the membership of this young system. 
It is interesting to note that the star HIP~86598 (F9V) at 72\,pc, formerly identified in $\beta$PMG
by \citet{2011kiss}, was recently proposed by \citet{2012song} to be a member
of the Scorpius-Centaurus region.

%%%%%%%%%%%%%%%%%%%%%%%%%%%%%%%%%%%%%%%%%%%%%%%%%%%%%%%%%%%%%%%%%%%%%%%%%%%%%%%%%%%%%%%%%
\section{Discussion} \label{chap:six}
%%%%%%%%%%%%%%%%%%%%%%%%%%%%%%%%%%%%%%%%%%%%%%%%%%%%%%%%%%%%%%%%%%%%%%%%%%%%%%%%%%%%%%%%%

Kinematics and Galactic positions alone are not sufficient to ascertain YMG membership, 
youth also needs to be established by verifying that candidates and {\it bona fide} members
share similar measurements. 
Here we discuss how stellar rotation and X-ray luminosity can be used for constraining the 
age of young low-mass stars.

%%%%%%%%%%%%%%%%%%%%%%%%%%%%%%%%%%%%%%%%%%%%%%%%%%%%%%%%%%%%%%%%%%%%%%%%%%%%%%%%%%%%%%%%%
\subsection{Rotation-Age Relation} \label{chap:sixun}
%%%%%%%%%%%%%%%%%%%%%%%%%%%%%%%%%%%%%%%%%%%%%%%%%%%%%%%%%%%%%%%%%%%%%%%%%%%%%%%%%%%%%%%%%

\begin{deluxetable}{lrrrrr}
\tabletypesize{\tiny}
\tablewidth{0pt}
\tablecolumns{6}
\tablecaption{$\log\,L_{\rm X}$ and \vsini\, average properties\label{tab:meanprop}}
\tablehead{
\colhead{Group of} & \colhead{$\log\,L_{\rm X}$} & \colhead{$\sigma_{\log\,L_{\rm X}}$} & \colhead{$\vsini$} & \colhead{$\sigma_{ \vsini}$} & \colhead{Number\tablenotemark{a}}\\
\colhead{stars} & \colhead{(erg s$^{-1}$)} & \colhead{(erg s$^{-1}$)} & \colhead{(km s$^{-1}$)} & \colhead{(km s$^{-1}$)} & \colhead{of stars} 
}
\startdata
\cutinhead{$\beta$PMG}
M0-M5 {\it Bona fide} & 29.6 & 0.1 & 8.0 & 1.4 & 7,7 \\
M0-M5 Candidate & 29.4 & 0.4 & 14.6 & 1.7 & 25,15 \\
M0-M2 Candidate & 29.5 & 0.2 & \nodata & \nodata & 10\\
M3-M5 Candidate & 29.3 & 0.5 & \nodata & \nodata & 15\\
\cutinhead{THA, COL, CAR and ARG}
M0-M5 {\it Bona fide} & \nodata & \nodata & \nodata & \nodata & 2 \\
M0-M5 Candidate & 29.4 & 0.3 & 22.5 & 2.5 & 42,31\\
\cutinhead{ABDMG}
M0-M5 {\it Bona fide} & 29.0 & 0.1 & 10.0 & 1.4 & 4,4 \\
M0-M5 Candidate & 29.1 & 0.2 & 16.0 & 1.9 & 17,14 \\
M0-M2 Candidate & 29.2 & 0.1 & \nodata & \nodata & 7\\
M3-M5 Candidate & 29.0 & 0.2 & \nodata & \nodata & 10\\
\cutinhead{Field}
M0-M5 & 27.5 & 0.6 & 3.1 & 1.4 & 39,39
\enddata
\tablenotetext{a}{Number of stars with $L_{\rm x}$ and $\vsini$ measurements, respectively.}
\end{deluxetable}

Past studies of the rotation periods of stars in young clusters and in the field have established
that stellar rotation is a function of mass, age and magnetic activity, and that the rotation rates
generally decrease with age \citep{2010soderblom}.  
The first parametrization of the empirical stellar rotation spin-down was introduced by \citet{1972skumanich},
using projected rotational velocity as a proxy of the rotation rate.
This parametrization takes the form of a power law, with an index of -0.5 (\vsini\ $\alpha$ to t$^{-1/2}$).
More recently, much attention has been paid to studying the stellar rotation rate evolution in the low-mass regime ($<$1 $\Msun$), which
eventually led to the development of gyrochronology \citep{2003barnes,2007barnes}.

One of the main outstanding problems within this area of research in understanding the 
angular momentum evolution of young late-type stars (K7V-M6V), for which the spin-up due to contraction competes with the spin-down due
to magnetic braking \citep{2012areiners}.
The stellar rotation rate evolution of M0V-M3V dwarfs can be broadly broken down into three main phases. 
During the first few million years, after the star is formed, the stellar rotation slows down due to interactions with the
circumstellar disk. 
After the primordial disk dissipates, the stellar rotation rate increases until the star reaches the zero-age main sequence, and then finally decreases
over time according to the power-law relation of \citet{1972skumanich}.
However, given the dispersion of rotation periods for old M dwarfs \citep{2011irwin}, the exact exponent of the power law may depend on the magnetic field 
geometry of the stars.
Observational data on the stellar rotation of low-mass stars at different ages and masses thus provide key 
insignts into their evolution \citep{2012areiners}.

To help investigate how stellar rotation evolves as a function of age, we compiled measurements of \vsini\, for the M0V-M5V 
{\it bona fide} members of YMGs (see Table~\ref{tab:bonafide_vsini}), which include 7 stars from $\beta$PMG, 4 stars from ABDMG and 2 stars for all THA, COL, CAR, and ARG YMGs. 
We excluded the binary systems from this analysis because their \vsini\ measurements could potentially be biased by blended line
profiles, for spectroscopic binaries, and because stellar rotation evolution in binary systems could be affected by angular momentum
exchanges between their components. 
We also compiled \vsini\ measurements for old M0V-M5V stars with a trigonometric parallax and
X-ray detection from several sources \citep{1998delfosse, 2005glebocki,2008reiners, 2009jenkins, 2013lepine}. 
This old sample comprises 39 stars and their properties are presented in Table~\ref{tab:x_mdwarf_old}.

\begin{figure}[!hbt]
\epsscale{1.3}
\plotone{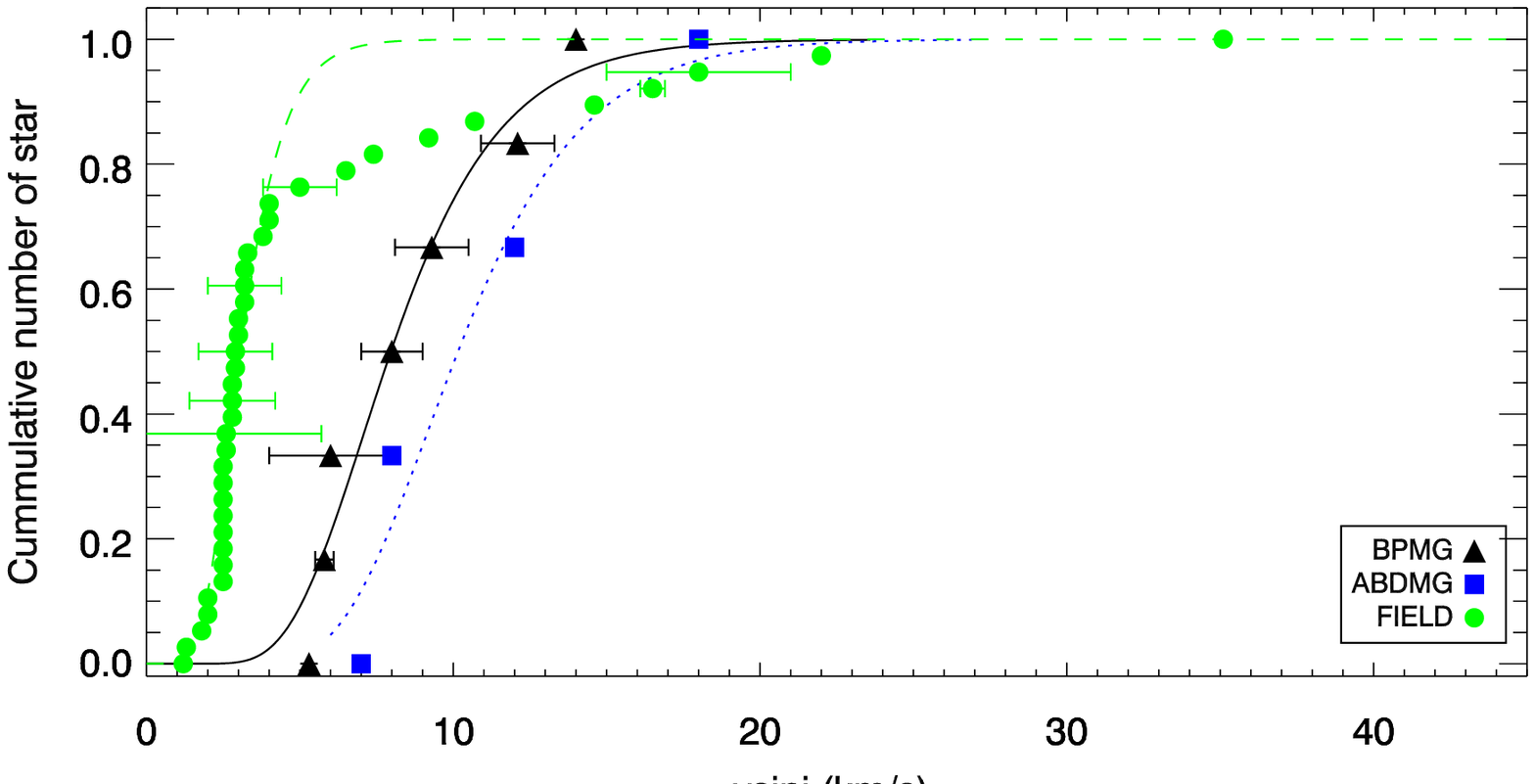}
\plotone{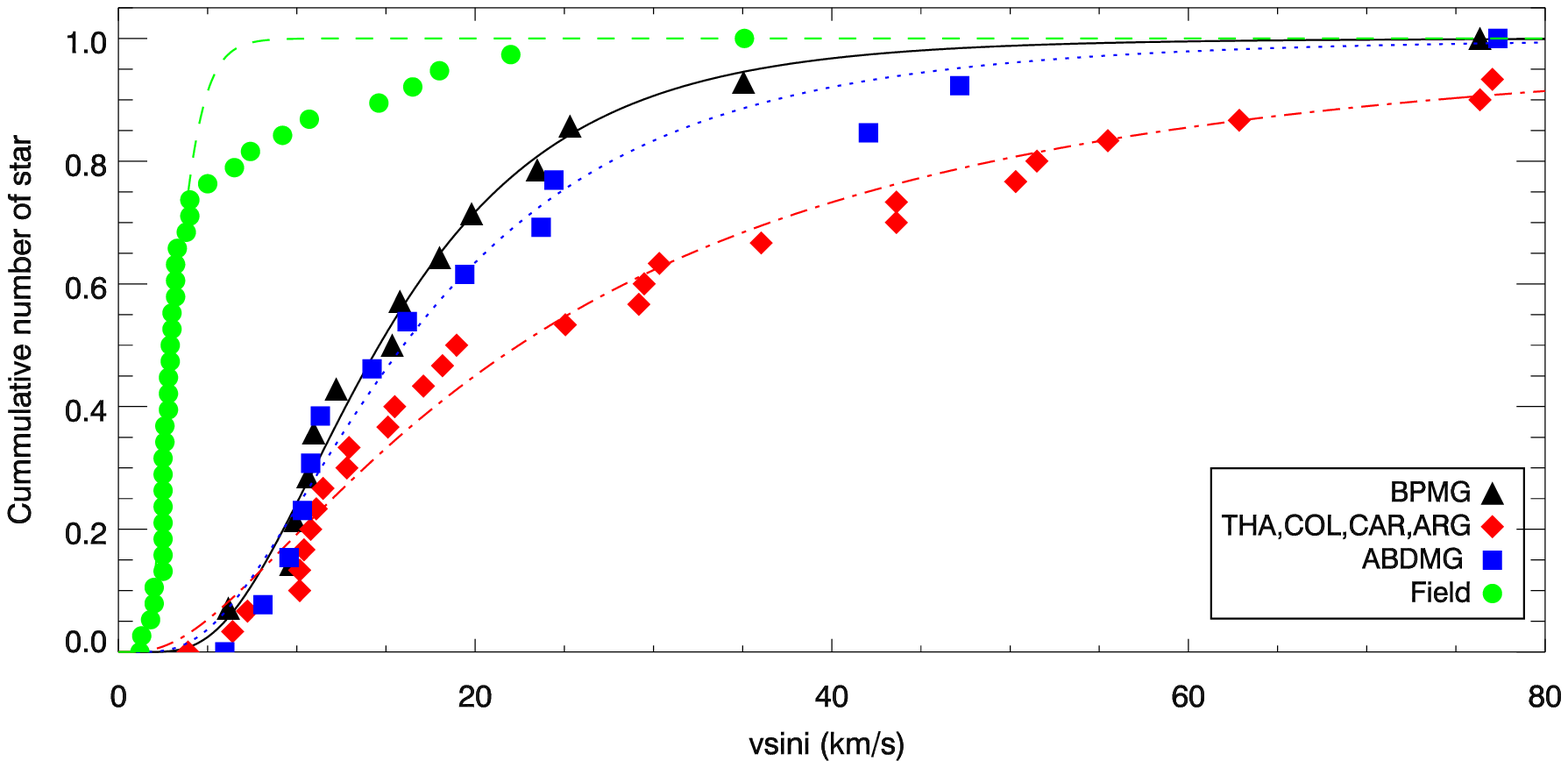}
\caption{\footnotesize{Top panel: \vsini\, cummulative distribution of M0V-M5V {\it bona fide} members excluding 
binaries (all kinds) compared to field distribution (filled green cercles).
The lines represent the adopted parameterization (see Table~\ref{tab:meanprop}).
Bottom panel: \vsini\, cummulative distribution of candidate members excluding binary systems.} \label{fig:fig8}}
\end{figure}

Figure~\ref{fig:fig8} (top) presents the cumulative distribution of \vsini\, for the M0V-M5V {\it bona fide} members of 
$\beta$PMG and ABDMG compared to that of the old sample.
We estimated the median and dispersion of these distributions by fitting a log-normal function, as proposed by \citet{2010weise}.
Table~\ref{tab:meanprop} summarises the median \vsini\, along with their dispersion for YMG and field stars. 
As visible in Figure~\ref{fig:fig8}, there is a trend for \vsini\, to decrease with age between $\beta$PMG and field stars.
We compared the distributions of \vsini\ pairwise by performing a double-sided Kolmogorov-Smirnov (K-S) test
using the $IDL$ routine $kstwo$.
We found that field \vsini\ measurements are clearly not drawn from the same distribution as that of $\beta$PMG; the K-S 
probability that both distributions are the same is only 0.1\%.
The bottom panel of the Figure~\ref{fig:fig8} presents the cumulative distributions of \vsini\, for 
our candidate members of the different YMGs with P$_{v} >$ 90\%, along with the distribution for the old sample.
Since our detection limits are between 3 and 8 \kms, depending on the slowly-rotating standard stars used, we reached only upper limits on 
\vsini\, for 28 candidates (see Table~\ref{tab:candprop}); these candidates were excluded from the figure.

To quantify the similarity between our samples of candidates and the {\it bona fide} members, we once again 
performed a K-S test as above.
Overall the candidate distributions are very similar to the {\it bona fide} YMG distributions with K-S probabilities of 20-73\%.
Accordingly, the probabilities that the \vsini\, of our candidates were drawn from the same distribution as those of the field sample are very low, 0.0002-0.0003\%.
Similar results were also found by \citet{2007scholz}.
It thus appears empirically that young low-mass stars in the age range defined by $\beta$PMG and ABDMG have larger \vsini\, than their old counterparts in the field.
In our candidates sample, there are 22 ultrafast rotators ($>$ 50 \kms), including 8 stars that were identified as binary systems; for
the 14 remainning fast rotators we cannot rule out the possibility that they are also unresolved binary systems.

\begin{deluxetable}{lrrrr}
\tabletypesize{\scriptsize}
\tablewidth{0pt}
\tablecolumns{5}
\tablecaption{Properties of {\it bona fide} members \label{tab:bonafide_vsini}}
\tablehead{
\colhead{Name of} & \colhead{$\vsini$\tablenotemark{a}}  & \colhead{$\pi$\tablenotemark{b}} & \colhead{$R_{\rm x}$} & \colhead{ $\log$ $L_{\rm x}$} \\
\colhead{star} & \colhead{(km s$^{-1}$)} & \colhead{(mas)} & \colhead{} & \colhead{(erg s$^{-1}$)}
}  
\startdata
\cutinhead{$\beta$PMG}
HIP 11152    & 6.0$\pm$2.0\tablenotemark{c}  &  34.86$\pm$2.84 & -2.85 &  29.58$\pm$0.08 \\
HIP 23200    & 14.0         &  38.64$\pm$2.54 & -3.15 &  29.58$\pm$0.01 \\
HIP 23309    & 5.8$\pm$0.3  &  37.34$\pm$1.13 & -3.37 &  29.40$\pm$0.08 \\
HIP 50156    & 8.0$\pm$1.0\tablenotemark{d}  &  43.32$\pm$1.80 & -3.35 &  29.31$\pm$0.03 \\
HIP 102409   & 9.3$\pm$1.2 &  100.91$\pm$1.06 & -2.98 &  29.75$\pm$0.01 \\
HIP 112312   & 12.1$\pm$1.2 &  42.84$\pm$3.61 & -2.68 &  29.69$\pm$0.03 \\
GJ 3331 A    & 5.3\tablenotemark{e}    & 51.98$\pm$1.30\tablenotemark{f}  & -2.91 &  29.58$\pm$0.04 \\
\cutinhead{ABDMG}
HIP 17695   & 18.0\tablenotemark{g} &  62.00$\pm$2.88 & -3.07 & 29.00$\pm$0.02 \\ 
HIP 31878   & 12.0\tablenotemark{g} &  44.74$\pm$0.91 & -3.70 & 28.84$\pm$0.05 \\
HIP 81084   & 7.0\tablenotemark{g}  &  32.60$\pm$2.47 & -3.40 & 28.98$\pm$0.04 \\
HIP 114066  & 8.0\tablenotemark{g}  &  40.81$\pm$1.60 & -3.00 & 29.42$\pm$0.01
\enddata
\tablenotetext{a}{Projected rotation velocity measurement from \citet{2006torres}, unless stated otherwise.}
\tablenotetext{b}{Parallax measurement from \citet{2007vanleeuwen}, unless stated otherwise.}
\tablenotetext{c}{\citet{2010schlieder}}
\tablenotetext{d}{\citet{2012herrero}}
\tablenotetext{e}{\citet{2012breiners}}
\tablenotetext{f}{\citet{2014riedel}}
\tablenotetext{g}{\citet{2009dasilva}}
\end{deluxetable}

%%%%%%%%%%%%%%%%%%%%%%%%%%%%%%%%%%%%%%%%%%%%%%%%%%%%%%%%%%%%%%%%%%%%%%%%%%%%%%%%%%%%%%%%%
\subsection{X-ray luminosity-Age Relation} \label{chap:sixdeux}
%%%%%%%%%%%%%%%%%%%%%%%%%%%%%%%%%%%%%%%%%%%%%%%%%%%%%%%%%%%%%%%%%%%%%%%%%%%%%%%%%%%%%%%%%

Past studies on stellar activity of low-mass stars have shown a saturation limit when the  
\vsini\, is compared to the $R_{\rm X}$ parameter, the ratio of X-ray to bolometric luminosity 
\citep{2005preibisch,2008mamajek}. 
This parameter is convenient as an activity proxy because it is independent of distance but 
it must rely on an estimate of the bolometric flux of the star.
In order to quantify the sensitivity of the $R_{\rm X}$ parameter vs age for YMG stars, we 
have compiled ROSAT X-ray luminosities
of all known {\it bona fide} low-mass members with spectral type later than M0V. 
This sample comprises 13 M stars.
The bolometric luminosities are determined using the $J$-band bolometric correction presented in 
\citet[][ see their Table 6, using $J-H$ index]{2013pecaut} for $\beta$PMG members and the $J$-band bolometric correction 
from \citet[][see their Equation 5]{2008casagrande} for ABDMG members and field dwarfs.
Again, all binary systems were excluded. 

\begin{figure}[!hbt]
\epsscale{1.2}
\plotone{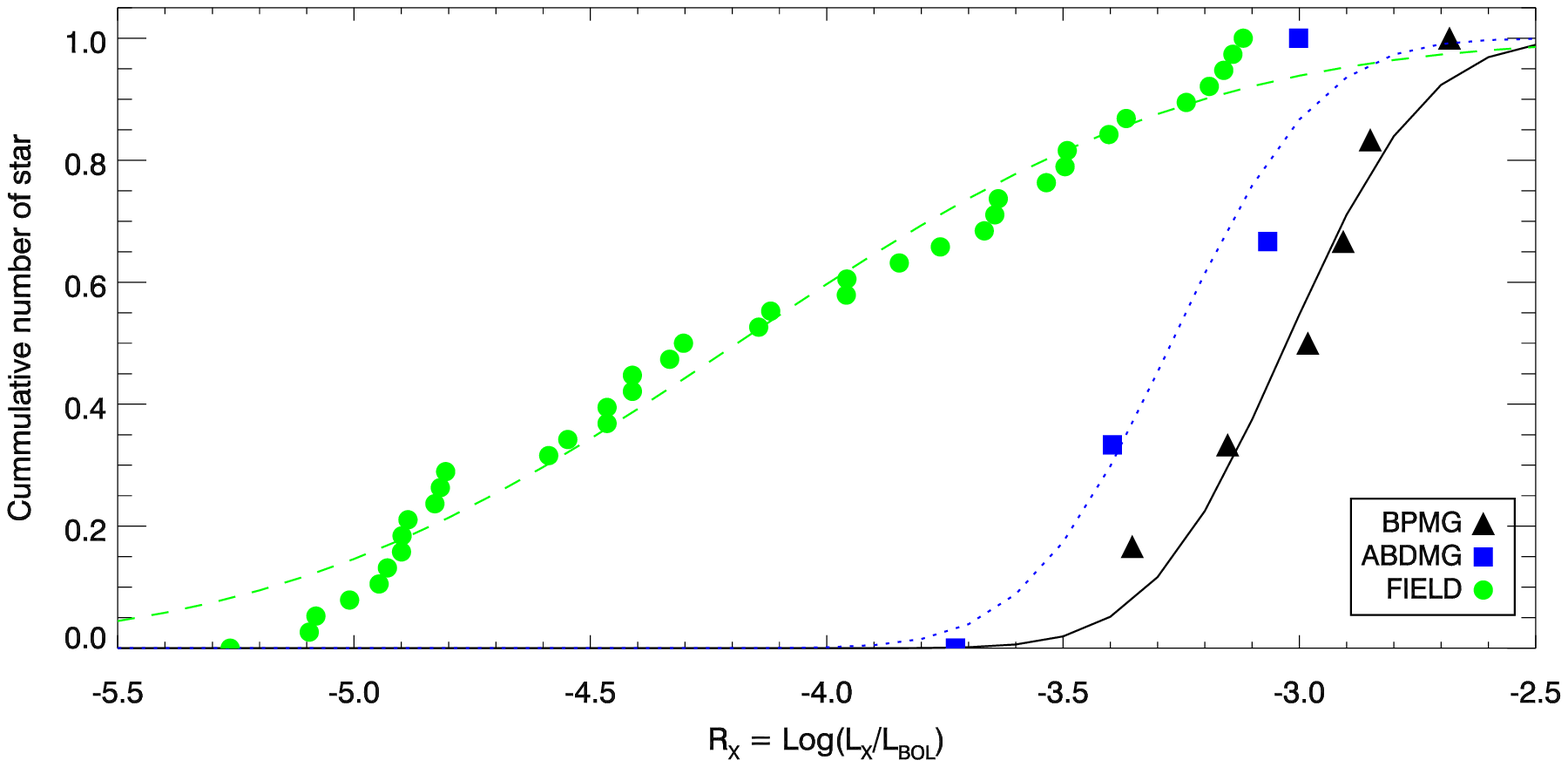}
\plotone{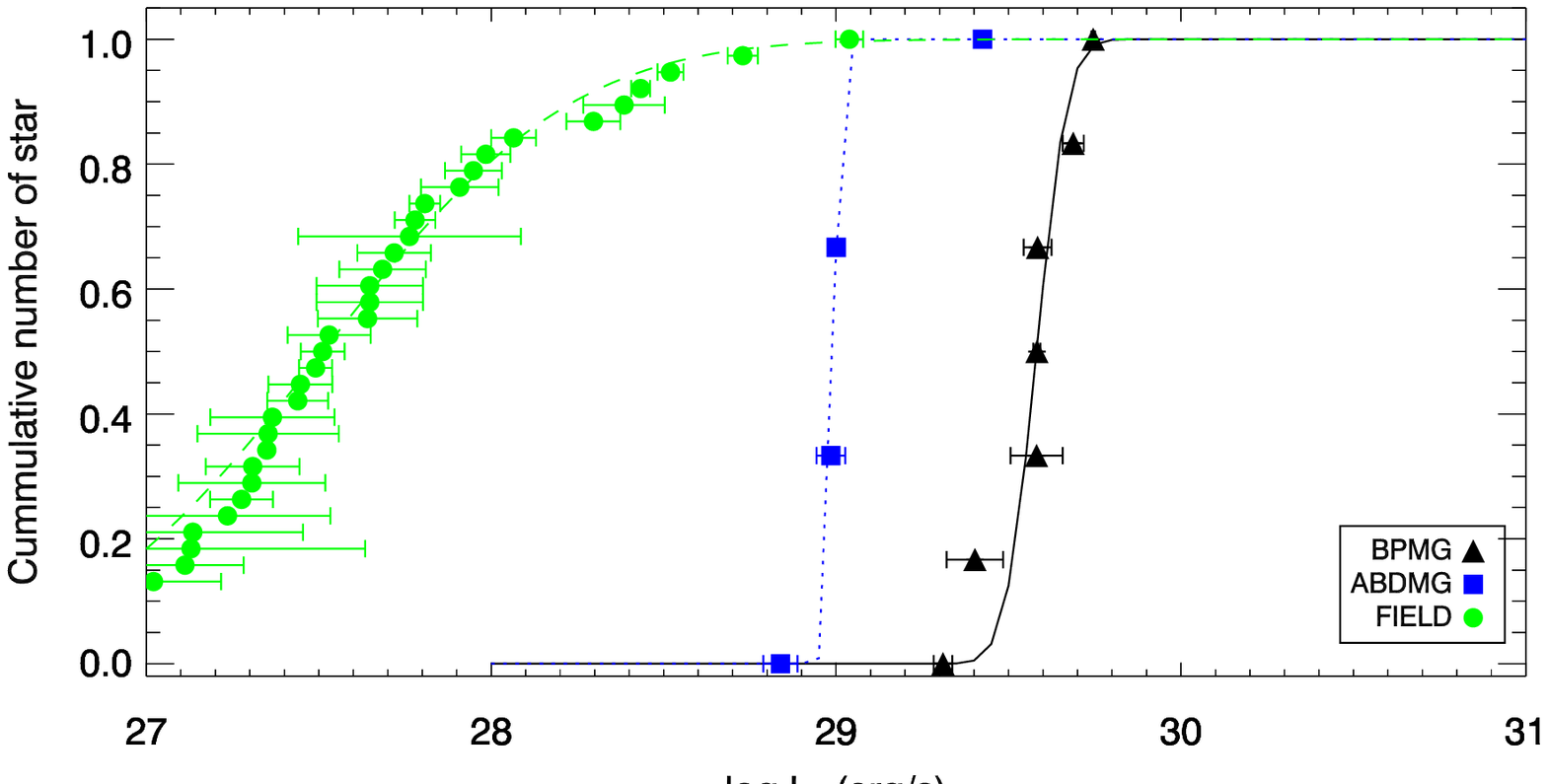}
\caption{\footnotesize{Top panel: Cummulative distribution of R$_{\rm X}$ ($\log~L_{\rm X}$/L$_{\rm bol}$) for 
M0V-M5V {\it bona fide} members, which exclude binaries.
Bottom panel: $\log~L_{\rm X}$ cummulative distribution for M0V-M5V {\it bona fide} members of $\beta$PMG (black triangles),
and ABDMG (blue squares).
} \label{fig:fig9}}
\end{figure}

The upper panel of Figure~\ref{fig:fig9} presents the cumulative distribution of 
$R_{\rm X}$ for young {\it bona fide} members and old field dwarfs.
While one can see a clear difference between young and old stars, there is no obvious distinction 
between $\beta$PMG and ABDMG members.

The bottom panel of Figure~\ref{fig:fig9} presents the cumulative distribution of 
$\log~L_{\rm X}$ for M0V-M5V {\it bona fide} members compared to the old low-mass population.
The X-ray luminosity is calculated using this formulae\footnote{http://heasarc.gsfc.nasa.gov/W3Browse/all/rassdsstar.html}: $L_{\rm x}$ = 1.2e$^{38}$~$f_{\rm x}$~$d_{\pi}^{2}$.
One clearly see a difference of $\sim$ 0.6\,dex between $\beta$PMG and ABDMG members.
Even if we take into account the uncertainties on the $ROSAT$ counts, HR1 ratio and parallax 
(error bar on Figure~\ref{fig:fig9}), the difference between the two young samples is significant.
A K-S test performed on both distributions shows that they are 
different at a confidence level of 94.4\%. 
We estimated the mean and dispersion of each distribution by fitting error functions (see 
the Table~\ref{tab:meanprop}).
We note that \citet{2005preibisch} found similar results for Chamaeleon association and the Pleiades group, where
these two groups have similar ages compared to $\beta$PMG and ABDMG.
Overall, relatively young stars represented by $\beta$PMG members are two orders of magnitude more luminous than 
M dwarfs in the field and this luminosity excess remains significant (a factor of $\sim$4) compared to older 
AMDMG members. 
This analysis shows that the X-ray luminosity is an excellent youth indicator for M dwarfs in the age range 
defined by $\beta$PMG and ABDMG members. 
This youth diagnostic is very complementary to the common method of using the presence of the lithium resonnance 
line at 6707.8 $\AA$. 
However, since lithium is rapidly depleted in M dwarfs, especially fully convective ones ($\sim$M3V), lithium is a 
useful youth indicator only for low-mass stars younger than a few 10$^{7}$ yrs. 
Using the X-ray luminosity as youth indicator extends by roughly an order of magnitude the age range covered 
by the lithium method.

\begin{figure}[!hbt]
\epsscale{1.2}
\plotone{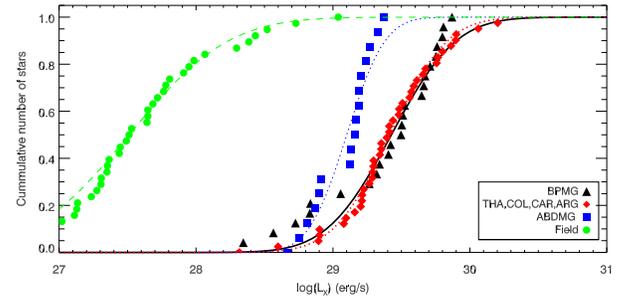}
\caption{\footnotesize{Cummulative distribution of $\log~L_{\rm X}^{s}$ for candidate members excluding 
binary systems compared to old field $\log~L_{\rm X}$ distribution.} \label{fig:fig6}}
\end{figure}

We performed the same analysis on our candidate members, where the X-ray luminosity was determined using 
the statistical distance (see Table~\ref{tab:candprop}), as a proxy of the parallax ($\log~L_{\rm X}^{s}$). 
As shown in Paper I, the statistical distance is a reliable estimate of the true distance within an uncertainty 
of $\sim$10\%.
Figure~\ref{fig:fig6} presents the cumulative distribution of $\log~L_{\rm X}^{s}$ for our
highly probable members compared to the low-mass field dwarfs. 
The amplitude of the error bars is $\sim$0.3~dex and take into account the uncertainties on both the $ROSAT$ flux 
and the statistical distance.
The same trend for the X-ray luminosity to increase at young ages is confirmed with the candidates. 
One can see in Figure~\ref{fig:fig6} that the $\beta$PMG distribution shows an excess of low-luminosity objects 
which could find an explanation if the statistical distance of these objects is somewhat different from reality. 
In fact, we do have a parallax for one $\beta$PMG candidate member and indeed, when one uses the true distance 
to infer its X-ray luminosity, its value is brought much closer to the nominal distribution for this association. 

\begin{figure}[!hbt]
\epsscale{1.2}
\plotone{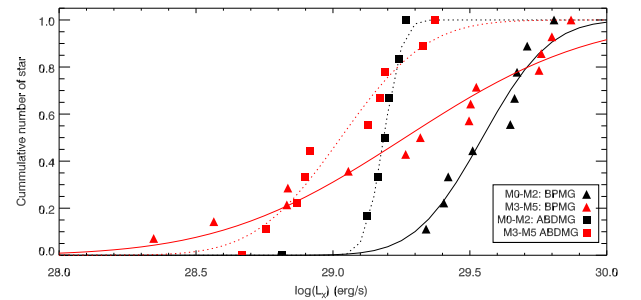}
\caption{\footnotesize{Cummulative distribution of $\log~L_{\rm X}$ for M0V-M2V and M3V-M5V of $\beta$PMG and ABDMG candidate members.
} \label{fig:fig18}}
\end{figure}

In order to investigate the behavior of the X-ray luminosity with spectral type (mass), we divide our candidates 
into two groups: M0V-M2V (partly convective stars) and M3V-M5V (fully convective stars). 
Figure~\ref{fig:fig18} presents the cumulative distribution of X-ray luminosity, for both groups in $\beta$PMG and ABDMG.
Table~\ref{tab:meanprop} summarises the median $\log~L_{\rm X}$ along with their dispersion for YMGs and field stars.
The X-ray luminosity for M3V-M5V $\beta$PMG candidates is slightly lower (with a higher dispersion) compared 
to the M0V-M2V group, and the same behavior is seen for ABDMG candidate members.
More observations are clearly needed to confirm this trend which may provide some insights for understanding the X-ray
property and magnetic activity of low-mass stars near the brown dwarf boundary.

%We note that in this age range and mass regime, stars have usually depleted their lithium abundance, therefore as shown in
%this section, X-ray luminosity should be used to better determined the age of low-mass stars.

%%%%%%%%%%%%%%%%%%%%%%%%%%%%%%%%%%%%%%%%%%%%%%%%%%%%%%%%%%%%%%%%%%%%%%%%%%%%%%%%%%%%%%%%%
\section{Summary and Conclusion} \label{chap:sept}
%%%%%%%%%%%%%%%%%%%%%%%%%%%%%%%%%%%%%%%%%%%%%%%%%%%%%%%%%%%%%%%%%%%%%%%%%%%%%%%%%%%%%%%%%

The study aims at extending the census of low-mass star members 
of seven young nearby associations.
A Bayesian statistical analysis described in Paper I was developed for identifying 
candidates based on a minimal set of observables (proper motion, photometry, position in the sky)
combined with a kinematic model of the association. 
This method yields a membership probability to a given association as well as the most probable 
(statistical) distance and the predicted radial velocity. 
In the present study, radial velocity measurements were used to better constrain the 
membership of the candidates.

Starting from a sample of 920 stars, all showing indicators of youth such as H$\alpha$ and/or
X-rays emission, our analysis selected 247 candidates with a membership probability over 90\%.
To confirm the predicted membership, we have secured radial velocity measurements for 202 
candidate members.
These measurements were combined with a compilation of 166 measurements from the literature to 
select a list of 130 young K and M stars with a membership probability to all YMGs exceeding 
90\% when the radial velocity is included in our Bayesian analysis.
A subgroup of 117 candidates are associated with a single YMG distributed as follows : 27 in 
$\beta$PMG, 22 in THA, 25 in COL, 7 in CAR, 18 in ARG and 18 in ABDMG.

We investigated the rotation-age relation and found that stellar rotation (\vsini) of young 
M dwarfs is significantly higher compared the their field counterparts. 
We also find that the X-ray luminosity of $\sim$12-22 Myr-old $\beta$PMG members are typically 
two orders of magnitude more luminous compared to field stars; this luminosity excess is 
a factor of $\sim$4 compared to older ($\sim$100 Myr) ABDMG members. 
Thus, the X-ray luminosity appears to be an excellent age discriminant for M dwarfs.
 
This work has unveiled a large population of highly probable low-mass members to nearby YMGs. 
A parallax, and ideally other indications of youth, are mandatory to firmly establish them as 
{\it bona fide} members of their respective association.

%We also see a hint for the \vsini\, of M3V-M5V ABDMG
%candidate members to be slightly higher compared to younger $\beta$PMG candidate members, a 
%trend expected theoretically for the evolution of young low-mass stars. 

%%%%%%%%%%%%%%%%%%%%%%%%%%%%%%%%%%%%%%%%%%%%%%%%%%%%%%%%%%%%%%%%%%%%%%%%%%%%%%%%%%%%%%%%%%%%%%%%%
\acknowledgments
%%%%%%%%%%%%%%%%%%%%%%%%%%%%%%%%%%%%%%%%%%%%%%%%%%%%%%%%%%%%%%%%%%%%%%%%%%%%%%%%%%%%%%%%%%%%%%%%%

The authors would like to thank Bernadette Rogers, German Gimeno, Michele Edwards, Elena Valenti and
the Gemini, CFHT and ESO staff for carrying out the observations. 
Special thanks to Anne-Marie Lagrange, Ansgar Reiners and Evgenya Shkolnik for interesting advices. 
Finally, we thank our referee for several comments which improved the quality of this paper.
 
This work was supported in part through grants from the 
the Fond de Recherche Qu\'eb\'ecois - Nature et Technologie and
the Natural Science and Engineering Research Council of Canada. 
This research has made use of the SIMBAD database, operated at 
Centre de Donn\'ees astronomiques de Strasbourg (CDS), 
Strasbourg, France. 
This research has made use of the VizieR catalogue access 
tool, CDS, Strasbourg, France \citep{2000ochsenbein}.

Based on observations obtained at the Gemini Observatory, which is operated by the
Association of Universities for Research in Astronomy, Inc., under a cooperative agreement
with the NSF on behalf of the Gemini partnership: the National Science Foundation (United
States), the Science and Technology Facilities Council (United Kingdom), the
National Research Council (Canada), CONICYT (Chile), the Australian Research Council
(Australia), Minist\'{e}rio da Ci\^{e}ncia, Tecnologia e Inova\c{c}\~{a}o (Brazil) 
and Ministerio de Ciencia, Tecnolog\'{i}a e Innovaci\'{o}n Productiva (Argentina).

The DENIS project has been partly funded by the SCIENCE and the HCM plans of
the European Commission under grants CT920791 and CT940627.
It is supported by INSU, MEN and CNRS in France, by the State of Baden-W\"urttemberg 
in Germany, by DGICYT in Spain, by CNR in Italy, by FFwFBWF in Austria, by FAPESP in Brazil,
by OTKA grants F-4239 and F-013990 in Hungary, and by the ESO C\&EE grant A-04-046.
Jean Claude Renault from IAP was the Project manager.  Observations were  
carried out thanks to the contribution of numerous students and young 
scientists from all involved institutes, under the supervision of  P. Fouqu\'e,  
survey astronomer resident in Chile.  

Funding for RAVE has been provided by: the Australian Astronomical Observatory; 
the Leibniz-Institut fuer Astrophysik Potsdam (AIP); the Australian National University; 
the Australian Research Council; the French National Research Agency; the German 
Research Foundation (SPP 1177 and SFB 881); the European Research Council (ERC-StG 240271 Galactica); 
the Istituto Nazionale di Astrofisica at Padova; The Johns Hopkins University; the National 
Science Foundation of the USA (AST-0908326); the W. M. Keck foundation; the Macquarie University; 
the Netherlands Research School for Astronomy; the Natural Sciences and Engineering Research 
Council of Canada; the Slovenian Research Agency; the Swiss National Science Foundation; 
the Science \& Technology Facilities Council of the UK; Opticon; Strasbourg Observatory; 
and the Universities of Groningen, Heidelberg and Sydney. The RAVE web site is at 
\url{http://www.rave-survey.org}. 

%%%%%%%%%%%%%%%%%%%%%%%%%%%%%%%%%%%%%%%%%%%%%%%%%%%%%%%%%%%%%%%%%%%%%%%%%%%%%%%%%%%%%%%%%
\bibliographystyle{apj}
\bibliography{references3}
%%%%%%%%%%%%%%%%%%%%%%%%%%%%%%%%%%%%%%%%%%%%%%%%%%%%%%%%%%%%%%%%%%%%%%%%%%%%%%%%%%%%%%%%%%%%%%%%

%%%%%%%%%%%%%%%%%%%%%%%%%%%%%%%%%%%%%%%%%%%%%%%%%%%%%%%%%%%%%%%%%%%%%%%%%%%%%%%%%%%%%%%%%
%\LongTables
%%%%%%%%%%%%%%%%%%%%%%%%%%%%%%%%%%%%%%%%%%%%%%%%%%%%%%%%%%%%%%%%%%%%%%%%%%%%%%%%%%%%%%%%%

%Properties of candidate members
\clearpage
\onecolumngrid
\LongTables
\begin{landscape}
\tabletypesize{\tiny}
\begin{deluxetable}{lrrrrrrrrrrrrrrrrr}
\tablewidth{0pt}
\tablecolumns{18}
\tablecaption{Properties of candidate members of young kinematic groups \label{tab:candprop}}
\tablehead{
\colhead{Name} & \colhead{$I_{c}$\tablenotemark{a}} & \colhead{$J$} & \colhead{$\mu_{\rm \alpha}$ cos $\delta$\tablenotemark{b}} & \colhead{$\mu_{\rm \delta}$\tablenotemark{b}} & \colhead{Spectral\tablenotemark{c}} & \colhead{X-ray\tablenotemark{d}} & \colhead{H$\alpha$\tablenotemark{e}} & \colhead{Li\tablenotemark{f}} & \colhead{vsini\tablenotemark{g}} & \colhead{RV$_{\rm meas}$\tablenotemark{h}} & \colhead{RV$_{\rm pred}$\tablenotemark{i}} & \colhead{$d_{\pi}$\tablenotemark{j}} & \colhead{$d_{s}$\tablenotemark{i}} & \colhead{P\tablenotemark{k}} & \colhead{Pv\tablenotemark{k}} & \colhead{log L$_{x}^{s}$\tablenotemark{d}} & \colhead{Refs.} \\
\colhead{(2MASS)} & \colhead{(mag)} & \colhead{(mag)} & \colhead{(mas yr$^{-1}$)} & \colhead{(mas yr$^{-1}$)} & \colhead{Type} & \colhead{log(f$_{x}$)} & \colhead{(\AA)} & \colhead{(m \AA)} & \colhead{(km s$^{-1}$)} & \colhead{(km s$^{-1}$)} & \colhead{(km s$^{-1}$)} & \colhead{(pc)} & \colhead{(pc)} & \colhead{(\%)} & \colhead{(\%)} & \colhead{(erg s$^{-1}$)} & \colhead{}
}
\startdata
\cutinhead{$\beta$ Pictoris moving group}
J00172353-6645124 & $ 9.94\pm0.05$ & $ 8.56\pm0.02$ & $  104.3\pm  1.0$ & $  -13.5\pm  1.0$ & M2.5V & $-11.77$ & $ 6.0$ &\ldots  & $<   7.4$\tablenotemark{zz} & $  10.7\pm 0.2$\tablenotemark{zz} & $  10.9\pm  1.5$ & $  39.0\pm  2.6$\tablenotemark{vv} & $  35\pm  3$ & $ 99.3$ & $ 99.9$ & $29.41\pm 0.19$ &       18  \\[0.5ex]
J00275023-3233060 & $10.29\pm0.03$\tablenotemark{vv} & $ 8.88\pm0.03$ & $   97.8\pm  1.1$ & $  -60.9\pm  4.0$ & M3.5Ve & $-11.59$ & $ 7.8$ &\ldots  & $   6.2\pm 1.8$\tablenotemark{zz} & $   8.8\pm 0.2$\tablenotemark{zz} & $   8.3\pm  1.4$ & $  32.3\pm  1.8$\tablenotemark{vv} & $  33\pm  2$ & $  1.8$ & $ 99.9$ & $29.53\pm 0.15$ &       17  \\[0.5ex]
J00275035-3233238 & $10.48\pm0.02$\tablenotemark{vv} & $ 8.97\pm0.03$ & $  117.4\pm  2.8$ & $  -29.3\pm  8.1$ & M3.5Ve & $-11.59$ & $ 5.3$ &\ldots  & $<   4.2$\tablenotemark{zz} & $   8.5\pm 0.2$\tablenotemark{zz} & $   8.3\pm  1.4$ & $  32.3\pm  1.8$\tablenotemark{vv} & $  32\pm  2$ & $ 97.2$ & $ 99.9$ & $29.51\pm 0.15$ &       17  \\[0.5ex]
J01132817-3821024 & $ 9.66\pm0.07$\tablenotemark{o} & $ 8.49\pm0.02$ & $  123.0\pm  1.1$ & $  -38.3\pm  1.2$ & M0.0+M1.0 & $-11.77$ & $ 2.5$ &\ldots  & $<   9.1$\tablenotemark{zz} & $  14.3\pm 0.5$\tablenotemark{zz} & $  11.9\pm  1.4$ & \ldots  & $  29\pm  2$ & $ 16.7$ & $ 99.3$ &\nodata &       18  \\[0.5ex]
J01365516-0647379 & $11.22\pm0.06$ & $ 9.71\pm0.02$ & $  175.1\pm  4.4$\tablenotemark{y} & $  -98.8\pm  4.4$\tablenotemark{y} & M4V+>L0 & $-12.03$ & $ 6.0$ &\ldots  & \ldots  & $  12.2\pm 0.4$\tablenotemark{uu} & $   9.5\pm  1.5$ & $  24.0\pm  0.4$\tablenotemark{uu} & $  21\pm  1$ & $ 28.3$ & $ 99.9$ &\nodata &       18  \\
\enddata
\end{deluxetable}
\clearpage
\end{landscape}
\twocolumngrid

%Probabilities of candidate members
\clearpage
\onecolumngrid
\LongTables
\begin{landscape}
\tabletypesize{\scriptsize}
\begin{deluxetable}{lccclccclccclccclccclccclccclccc}
\tablewidth{640pt}
\tablecolumns{32}
\tablecaption{Membership probabilities of all candidates\tablenotemark{a} \label{tab:allcand}}
\tablehead{
\colhead{} & \multicolumn{3}{c}{$\beta$PMG} & \colhead{} & \multicolumn{3}{c}{TWA} & \colhead{} & \multicolumn{3}{c}{THA} & \colhead{} & \multicolumn{3}{c}{COL} &  & \multicolumn{3}{c}{CAR} & \colhead{} & \multicolumn{3}{c}{ARG} & \colhead{} & \multicolumn{3}{c}{ABDMG} & \colhead{} & \multicolumn{3}{c}{Field} \\
\cline{2-4}
\cline{6-8}
\cline{10-12}
\cline{14-16}
\cline{18-20}
\cline{22-24}
\cline{26-28}
\cline{30-32}
\colhead{Name} & \colhead{P} & \colhead{P$_{v}$} & \colhead{P$_{v+\pi}$} & \colhead{} & \colhead{P} & \colhead{P$_{v}$} & \colhead{P$_{v+\pi}$} & \colhead{} & \colhead{P} & \colhead{P$_{v}$} & \colhead{P$_{v+\pi}$} & \colhead{} & \colhead{P} & \colhead{P$_{v}$} & \colhead{P$_{v+\pi}$} & \colhead{} & \colhead{P} & \colhead{P$_{v}$} & \colhead{P$_{v+\pi}$} & \colhead{} & \colhead{P} & \colhead{P$_{v}$} & \colhead{P$_{v+\pi}$} & \colhead{} & \colhead{P} & \colhead{P$_{v}$} & \colhead{P$_{v+\pi}$} & \colhead{} & \colhead{P} & \colhead{P$_{v}$} & \colhead{P$_{v+\pi}$}
}
\startdata
J00172353-6645124 & $ 99.3$ & $ 99.9$ & $ 99.9$ & $ $ & $  0.0$ & $  0.0$ & $  0.0$ & $ $ & $  0.0$ & $  0.0$ & $  0.0$ & $ $ & $  0.0$ & $  0.0$ & $  0.0$ & $ $ & $  0.0$ & $  0.0$ & $  0.0$ & $ $ & $  0.5$ & $  0.0$ & $  0.0$ & $ $ & $  0.1$ & $  0.0$ & $  0.0$ & $ $ & $  0.0$ & $  0.0$ & $  0.0$ \\[0.5ex]
J00275023-3233060 & $  1.8$ & $ 99.9$ & $ 99.9$ & $ $ & $  0.0$ & $  0.0$ & $  0.0$ & $ $ & $ 94.6$\tablenotemark{b} & $  0.0$ & $  0.0$ & $ $ & $  3.6$\tablenotemark{b} & $  0.0$ & $  0.0$ & $ $ & $  0.0$ & $  0.0$ & $  0.0$ & $ $ & $  0.0$ & $  0.0$ & $  0.0$ & $ $ & $  0.1$ & $  0.2$ & $  0.0$ & $ $ & $  0.0$ & $  0.2$ & $  0.0$ \\[0.5ex]
J00275035-3233238 & $ 97.2$ & $ 99.9$ & $ 99.9$ & $ $ & $  0.0$ & $  0.0$ & $  0.0$ & $ $ & $  2.2$\tablenotemark{b} & $  0.0$ & $  0.0$ & $ $ & $  0.3$ & $  0.0$ & $  0.0$ & $ $ & $  0.0$ & $  0.0$ & $  0.0$ & $ $ & $  0.1$ & $  0.0$ & $  0.0$ & $ $ & $  0.0$ & $  0.0$ & $  0.0$ & $ $ & $  0.1$ & $  0.0$ & $  0.0$ \\[0.5ex]
J00281434-3227556 & $ 81.1$ & $ 83.6$ & \ldots  & $ $ & $  0.0$ & $  0.0$ & \ldots  & $ $ & $ 16.4$\tablenotemark{b} & $ 12.0$\tablenotemark{b} & \ldots  & $ $ & $  2.4$\tablenotemark{b} & $  4.4$\tablenotemark{b} & \ldots  & $ $ & $  0.0$ & $  0.0$ & \ldots  & $ $ & $  0.0$ & $  0.0$ & \ldots  & $ $ & $  0.0$ & $  0.0$ & \ldots  & $ $ & $  0.1$ & $  0.1$ & \ldots  \\[0.5ex]
J00340843+2523498 & $  0.0$ & $  0.0$ & \ldots  & $ $ & $  0.0$ & $  0.0$ & \ldots  & $ $ & $  0.0$ & $  0.0$ & \ldots  & $ $ & $  0.0$ & $  0.0$ & \ldots  & $ $ & $  0.0$ & $  0.0$ & \ldots  & $ $ & $  0.0$ & $  0.0$ & \ldots  & $ $ & $ 99.4$ & $ 99.9$ & \ldots  & $ $ & $  0.6$ & $  0.1$ & \ldots  \\
\enddata
\tablenotetext{a}{Membership probability (P), membership probability including radial velocity information (P$_{v}$), 
or membership probability including radial velocity and parallax informations (P$_{v+\pi}$).}
\tablenotetext{b}{Membership probability (P or P$_{v}$ or P$_{v+\pi}$) for which the binary hypothesis has a higher probability.}
\end{deluxetable}
\clearpage
\end{landscape}
\twocolumngrid

\LongTables
\tabletypesize{\scriptsize}
\twocolumngrid
\begin{deluxetable}{lcccc}
\tablewidth{0pt}
\tablecolumns{5}
\tablecaption{Individual radial and projected rotational velocity measurements \label{tab:indivi}}
\tablehead{
\colhead{Name} & \colhead{Date} & \colhead{Instrument} & \colhead{RV} & \colhead{vsini} \\
\colhead{of star} & \colhead{(DDMMYY)} & \colhead{} & \colhead{(kms$^{-1}$)} & \colhead{(kms$^{-1}$)}
}
\startdata
J00240899-6211042 & $081109$ & $ PHOENIX$ & $    5.8\pm  0.7$ & $   20.7\pm  2.0$ \\[0.5ex]
J00172353-6645124 & $260809$ & $ PHOENIX$ & $   10.8\pm  0.2$ & $    6.3\pm  1.8$ \\[0.5ex]
J00172353-6645124 & $240910$ & $ PHOENIX$ & $   10.0\pm  0.5$ & $<    8.7$ \\[0.5ex]
J00171443-7032021 & $010909$ & $ PHOENIX$ & $   -4.5\pm  0.4$ & $   10.1\pm  3.5$ \\[0.5ex]
J00171443-7032021 & $170910$ & $ PHOENIX$ & $   -2.5\pm  0.6$ & $   12.2\pm  2.6$ \\[0.5ex]
J00233468+2014282ab & $290513$ & $ESPADONS$ & $   -1.6\pm  0.2$ & $<    4.6$ \\[0.5ex]
J00233468+2014282ab & $310513$ & $ESPADONS$ & $   -1.7\pm  0.2$ & $<    4.2$ \\[0.5ex]
J00275023-3233060a & $210713$ & $  CRIRES$ & $    8.8\pm  0.2$ & $    6.2\pm  1.8$ \\
\enddata
\end{deluxetable}
\onecolumngrid

% Compilation RV
\clearpage
\twocolumngrid
\LongTables
%\begin{landscape}
\tabletypesize{\scriptsize}
\begin{deluxetable}{lcccc}
\tablewidth{640pt}
\tablecolumns{5}
\tablecaption{Compilation of RV from the literature \label{tab:compilrv}}
\tablehead{
\colhead{Name} & \colhead{Spectral} & \colhead{RV} & \colhead{RV} & \colhead{RV} \\
\colhead{2MASS} & \colhead{Type} & \colhead{(km s$^{-1}$)} & \colhead{(km s$^{-1}$)} & \colhead{(km s$^{-1}$)}
}
\startdata
J00340843+2523498 & K7Ve* & $  -12.4\pm  2.0$\tablenotemark{c} & $   -8.8\pm  0.2$\tablenotemark{a} &\ldots  \\[0.5ex]
J01220441-3337036 & K7Ve & $    4.8\pm  0.8$\tablenotemark{e} & $    3.0\pm  1.4$\tablenotemark{c} & $    4.7\pm  0.3$\tablenotemark{a} \\[0.5ex]
J01225093-2439505 & M3.5V & $    9.6\pm  0.7$\tablenotemark{g} & $   11.4\pm  0.2$\tablenotemark{a} &\ldots  \\[0.5ex]
J01521830-5950168 & M2-3V & $    8.5\pm  2.3$\tablenotemark{h} & $    7.9\pm  1.6$\tablenotemark{d} &\ldots  \\[0.5ex]
J02303239-4342232 & K5Ve & $   16.3\pm  1.1$\tablenotemark{e} & $   14.5\pm  2.4$\tablenotemark{h} &\ldots  \\[0.5ex]
J02365171-5203036 & M2Ve & $   16.0\pm  0.1$\tablenotemark{e} & $   14.6\pm  2.1$\tablenotemark{a} &\ldots  \\[0.5ex]
J03494535-6730350 & K7V & $   16.3\pm  3.2$\tablenotemark{h} & $   16.8\pm  0.2$\tablenotemark{a} &\ldots  \\[0.5ex]
J04435686+3723033 & M3Ve & $    6.2\pm  2.0$\tablenotemark{c} & $    6.4\pm  0.2$\tablenotemark{a} &\ldots  \\
\enddata
\tablecomments{(a) this work; (b) \citealp{2013binks} (c) \citealp{2010schlieder}; (d) \citealp{2011kiss}; 
(e) \citealp{2006torres}; (f) \citealp{2012bowler}; (g) \citealp{2013bowler}; 
(h) \citealp[RAVE;][]{2008zwitter}; (i) \citealp{2012shkolnik};  (j) \citealp{2013moor};
(k) \citealp{2001montes}; (l) \citealp{2011rodriguez}. } 
\end{deluxetable}
\clearpage
%\end{landscape}
\twocolumngrid

%Compilation vsini
\clearpage
\twocolumngrid
\LongTables
%\begin{landscape}
\tabletypesize{\scriptsize}
\begin{deluxetable}{lccc}
\tablewidth{640pt}
\tablecolumns{4}
\tablecaption{Compilation of \vsini\, from the literature \label{tab:compilvsini}}
\tablehead{
\colhead{Name} & \colhead{Spectral} & \colhead{\vsini} & \colhead{\vsini} \\
\colhead{2MASS} & \colhead{Type} & \colhead{(km s$^{-1}$)} & \colhead{(km s$^{-1}$)}
}
\startdata
J01220441-3337036 & K7Ve & $    4.9\pm  1.2$\tablenotemark{b} & $<    5.8$\tablenotemark{a} \\[0.5ex]
J02365171-5203036 & M2Ve & $   36.0\pm  3.6$\tablenotemark{b} & $   38.1\pm  3.5$\tablenotemark{a} \\[0.5ex]
J02414683-5259523 & K6Ve & $   80.4\pm  3.8$\tablenotemark{b} & $   80.0$\tablenotemark{c} \\[0.5ex]
J02414730-5259306 & M2.5V & $   10.4\pm  1.9$\tablenotemark{b} & $    8.0$\tablenotemark{c} \\[0.5ex]
J04480066-5041255 & K7Ve(vb) & $    5.4\pm  1.2$\tablenotemark{b} & $<    5.9$\tablenotemark{a} \\[0.5ex]
J05332558-5117131 & K7V & $   18.4\pm  1.2$\tablenotemark{b} & $<    6.2$\tablenotemark{a} \\[0.5ex]
J11102788-3731520 & M4Ve+M4Ve & $   11.6\pm  1.2$\tablenotemark{b} & $<   12.4$\tablenotemark{a} \\[0.5ex]
J11211723-3446454 & M1Ve & $   12.3\pm  1.2$\tablenotemark{b} & $   10.0$\tablenotemark{c} \\[0.5ex]
J11211745-3446497 & M1Ve & $   12.0\pm  1.2$\tablenotemark{b} & $   10.0$\tablenotemark{c} \\[0.5ex]
J17130733-8552105 & M0V & $   85.0$\tablenotemark{c} & $   99.2\pm  9.3$\tablenotemark{a} \\[0.5ex]
J21100535-1919573 & M2V & $    5.0$\tablenotemark{c} & $    9.7\pm  1.2$\tablenotemark{a} \\[0.5ex]
J23261069-7323498 & M0V(sb) & $   61.0\pm  0.8$\tablenotemark{b} & $   64.1\pm  8.6$\tablenotemark{a} \\[0.5ex]
J23323085-1215513 & M0Ve & $    8.8\pm  1.2$\tablenotemark{b} & $<    9.0$\tablenotemark{a} \\[0.5ex]
\enddata
\tablecomments{(a) this work; (b) \citealp{2006torres}; (c) \citealp{2005glebocki}. }
\end{deluxetable}
\clearpage
%\end{landscape}
\twocolumngrid

%Compilation old M dwarfs
\clearpage
\twocolumngrid
\LongTables
%\begin{landscape}
\tabletypesize{\scriptsize}
\begin{deluxetable}{lcccccc}
\tablewidth{0pt}
\tablecolumns{7}
\tablecaption{Compilation of X-ray emission from field M dwarfs \label{tab:x_mdwarf_old}}
\tablehead{
\colhead{Name} & \colhead{Spectral} & \colhead{$\log f_{\rm x}$\tablenotemark{a}} & \colhead{$\pi$} & \colhead{$\log L_{\rm x}$\tablenotemark{b}} & \colhead{R$_{\rm x}$} & \colhead{vsini} \\
\colhead{2MASS} & \colhead{Type} & \colhead{(erg s$^{-1}$ cm$^{-2}$)} & \colhead{(mas)} & \colhead{(erg s$^{-1}$)} & \colhead{ } & \colhead{(km s$^{-1}$)}
}
\startdata
J01123052-1659570 & M4.5V & $ -11.73\pm 0.07$ & $ 269.10\pm 7.60$\tablenotemark{c} & $  27.49\pm 0.05$ & $  -3.54$ & $   2.50$\tablenotemark{i} \\[0.5ex]
J02001278+1303112 & M4.5V & $ -11.94\pm 0.11$ & $ 222.00\pm 5.00$\tablenotemark{d} & $  27.45\pm 0.09$ & $  -3.64$ & $   3.80$\tablenotemark{j} \\[0.5ex]
J02164119-3059181 & M3V & $ -11.87\pm 0.08$ & $  69.81\pm 3.10$\tablenotemark{c} & $  28.52\pm 0.04$ & $  -3.37$ & $   5.00$\tablenotemark{k} \\[0.5ex]
J02441537+2531249 & M3V & $ -12.52\pm 0.23$ & $ 133.16\pm 2.26$\tablenotemark{c} & $  27.31\pm 0.21$ & $  -4.55$ & $   2.80$\tablenotemark{g} \\[0.5ex]
J05015746-0656459 & M4V & $ -12.31\pm 0.14$ & $ 163.00\pm26.00$\tablenotemark{e} & $  27.35\pm 0.00$ & $  -3.96$ & $   3.20$\tablenotemark{e} \\[0.5ex]
J06000351+0242236 & M4V & $ -11.53\pm 0.08$ & $ 190.90\pm 1.90$\tablenotemark{f} & $  27.98\pm 0.07$ & $  -3.49$ & $   7.40$\tablenotemark{g} \\[0.5ex]
J10112218+4927153 & M0V & $ -12.02\pm 0.09$ & $ 205.21\pm 0.54$\tablenotemark{c} & $  27.44\pm 0.09$ & $  -5.26$ & $   2.80$\tablenotemark{j} \\[0.5ex]
J10121768-0344441 & M1.5V & $ -12.51\pm 0.19$ & $ 127.08\pm 1.90$\tablenotemark{c} & $  27.37\pm 0.18$ & $  -4.90$ & $   1.80$\tablenotemark{i} \\
\enddata
\tablenotetext{a}{ROSAT X-ray flux calculated from \citet[][Section 5.4]{2006riaz} in erg s$^{-1}$ cm$^{-2}$.}
\tablenotetext{b}{ROSAT X-ray luminosity calculated from formulae on http://heasarc.gsfc.nasa.gov/W3Browse/all/rassdsstar.html. }
\tablecomments{(c) \citet{2007vanleeuwen}; (d) \citet{2012houdebine}; (e) \citet{1998delfosse}; (f) \citet{2011lepine}, (g) \citet{2009jenkins}; 
(h) \citet{2005lepine}; (i) \citet{2012breiners}; (j) \citet{2005glebocki}; (k) \citet{2006torres} }
\end{deluxetable}
\clearpage
%\end{landscape}
\twocolumngrid

\end{document}